\documentclass[%
 reprint,
 superscriptaddress,
 amsmath,amssymb,
 aps,
 pre,
 onecolumn,
]{revtex4-1}
\usepackage{graphicx}% Include figure files
\usepackage{dcolumn}% Align table columns on decimal point
\usepackage{bm}% bold math
\usepackage{upgreek}
\usepackage{amsmath}
\usepackage{xcolor}

\begin{document}

\title{Transition from contact to hydrodynamic lubrication over rough surfaces}

\author{Vincent Bertin}
\email[]{vincent.bertin.1@univ-amu.fr}
\affiliation{Aix Marseille University, CNRS, IUSTI UMR 7343, Marseille 13453, France}
\affiliation{Universit\'{e} C\^{o}te d'Azur, CNRS, INPHYNI UMR 7010, Nice 06200, France}

\author{ Olivier Pouliquen}
\affiliation{Aix Marseille University, CNRS, IUSTI UMR 7343, Marseille 13453, France}

\date{\today}% It is always \today, today,

\begin{abstract}
\textbf{Abstract:} We present a theoretical analysis of frictional transitions along the Stribeck curve for rough elastic contacts lubricated by a Newtonian fluid. Building on the mean-field framework of Persson and Scaraggi (J. Phys.: Condens. Matter \textbf{21} (2009) 185002), we formulate a minimal elastohydrodynamic model that couples contact mechanics and lubrication through a homogenized pressure decomposition. Dimensional analysis reveals three independent dimensionless parameters governing the frictional response, which correspond to a dimensionless speed, normal load, and surface roughness.

Using asymptotic expansions, we first characterize the boundary and hydrodynamic lubrication regimes, which arise naturally as the quasistatic and smooth-surface limits of the model. In both limits, the contact morphology converges toward Hertzian contact in the regime of large elastic deformation, with boundary layers regularizing the separation profile at the edge of the contact zone. 

We then analyze the mixed lubrication regime and derive asymptotic expressions for the friction coefficient in the low- and high-speed limits. At high speeds, friction decomposes into a viscous contribution and a residual contact term, leading to a roughness- and load-dependent criterion for the transition to hydrodynamic lubrication that departs from constant-$\Lambda$ ratio theories. At low speeds, friction reduction results from the progressive redistribution of the applied load between asperity contact and hydrodynamic pressure, yielding a characteristic transition speed from boundary to mixed lubrication. These results are summarized in a phase diagram that generalizes the classical Stribeck curve to a multidimensional parameter space.

\end{abstract}

\maketitle

\section{Introduction}

Lubrication is one of the most common and effective strategies for reducing sliding friction and wear between solid surfaces~\cite{persson2013sliding}. It plays a central role across a wide spectrum of applications: from the operation of large‐scale mechanical systems such as gears, bearings, and engines, to microelectromechanical devices~\cite{bhushan2001tribology}, biomedical implants~\cite{moghadasi2022review}, and even natural systems such as synovial joints~\cite{jahn2016lubrication}. A central challenge across these systems is to ensure robust lubrication so that the surfaces remain separated and direct contact is avoided.

\begin{figure}[h!]
\centerline{\includegraphics{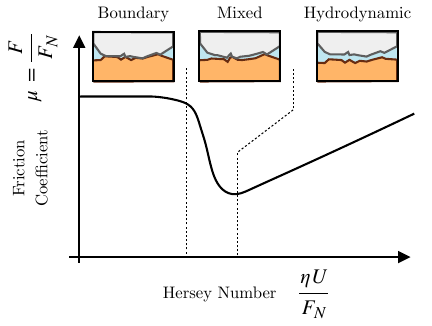}}	
\caption{\label{fig1} \textbf{Stribeck curve schematic.} Friction coefficient, ratio between shear force $F$ over the normal force per unit length $F_N$, versus the Hersey Number $\eta U/F_N$, where $\eta$ and $U$ are the lubricant viscosity and sliding speed respectively. Three regimes of boundary, mixed and hydrodynamic lubrication are encountered as the Hersey number increases.  }
\end{figure}

The Stribeck curve, illustrated in Fig.~\ref{fig1}, provides a qualitative framework for characterizing the frictional response under lubrication. It represents the coefficient of friction $\mu$, ratio of tangential to normal forces, as a function of the Hersey number, defined as the product of viscosity and sliding velocity divided by the applied normal load per unit length. Three regimes are typically distinguished:
(i) boundary lubrication, occurring at high loads and low speeds, where the lubricant is largely expelled and friction and wear are governed by direct solid–solid contact;
(ii) hydrodynamic or elastohydrodynamic lubrication (EHL), at high speeds and low loads, where a liquid film is entrained and supports the load, with friction dominated by viscous shear; and
(iii) an intermediate mixed lubrication regime, where both asperity contact and hydrodynamic pressure contribute significantly, and the friction coefficient decreases with speed. Although the existence of these three regimes is well established, the mechanisms governing the transitions between them remain poorly understood.

The boundary-to-mixed transition has received little attention in the literature. Classical contact mechanics models, such as the Greenwood–Williamson asperity model~\cite{greenwood1966contact}, and more advanced statistical theories of roughness such as Persson’s model~\cite{persson2001theory}, provide detailed descriptions of the stochastic mechanics of asperity contact. Both frameworks predict that the real area of contact increases linearly with the applied normal load, thereby providing a microscopic basis for Coulomb’s friction law. However, these approaches neglect the role of fluid flow, and thus cannot capture the coupled interplay between solid contact and hydrodynamic pressure that governs the onset of mixed lubrication. The aim of this article is to fill this gap. 
It is worth mentioning that the frictional response of soft materials such as rubbers or gels often departs from Coulomb’s law, being governed instead by additional physico-chemical processes~\cite{gong2006friction}, such as chain adsorption~\cite{schallamach1963theory,persson2025rubber}, viscoelastic losses~\cite{grosch1963relation,persson2001theory}, or poroelastic dissipation~\cite{delavoipiere2018friction,cuccia2020pore}. In this article, we leave those effects aside and will restrict ourselves to elastic interfaces. 

The mixed-to-hydrodynamic transition has traditionally been analyzed through extensions of elastohydrodynamic lubrication (EHL) theory. Classical EHL provides a description of pressure generation and hydrodynamic friction between nominally smooth, conforming surfaces, and thus applies directly in the hydrodynamic regime, where roughness $\ll$ film thickness~\cite{venner2000multi,lugt2011review,greenwood2020elastohydrodynamic}. However, it cannot easily accommodate the stochastic nature of real topographies. A central difficulty is the multiscale character of rough surfaces, which typically display correlated roughness spanning several orders of magnitude in length scale~\cite{jacobs2017quantitative,ardah2025comprehensive}. A major advance was the homogenization framework of Patir and Cheng~\cite{patir1978,patir1979application,Tripp1983}, which introduced so-called flow factors to account for the averaged influence of surface roughness on the Reynolds equation. These models predict that roughness significantly alters the pressure distribution once the mean film thickness becomes comparable to a few times the root-mean-square (rms) roughness, $h_\mathrm{rms}$. This observation motivated the introduction of the dimensionless $\Lambda$ ratio, defined as the ratio of central film thickness to $h_\mathrm{rms}$. It is often argued that the mixed-to-hydrodynamic transition is controlled by this thickness ratio $\Lambda = 3$~\cite{spikes1997mixed}. 

Nevertheless, experimental studies report a much broader range of transition values, from $\Lambda \simeq 0.7–11$~\cite{bongaerts2007soft}, $\Lambda \simeq 0.7–1.2$ \cite{petrova2019fluorescence}, $\Lambda \simeq 0.9–2.3$ \cite{sadowski2019friction}, and $\Lambda \simeq 16–2{,}736$ \cite{dong2023transition}. 
This broad scatter suggests that there is no universal transition scenario governed solely by the ratio of film thickness to roughness. Instead, different physical mechanisms may trigger the onset of contact depending on the system under consideration.
Indeed, in many real situations the transition is initiated or strongly influenced by physico-chemical mechanisms, such as long-range adhesive forces leading to “jump-to-contact”~\cite{dong2025role}, mechanical surface instabilities including wrinkling or Schallamach waves~\cite{dong2023transition,wu2007stick}, or viscoelastic stresses in lubricant film~\cite{oratis2025viscoelastic}. This broad scatter emphasizes that a global framework of mixed lubrication is still lacking -- one that can capture the interplay of contact mechanics, fluid dynamics, and physico-chemical effects, and thereby clarify the key parameters and scaling laws governing transitions.

\begin{figure}[h!]
\centerline{\includegraphics{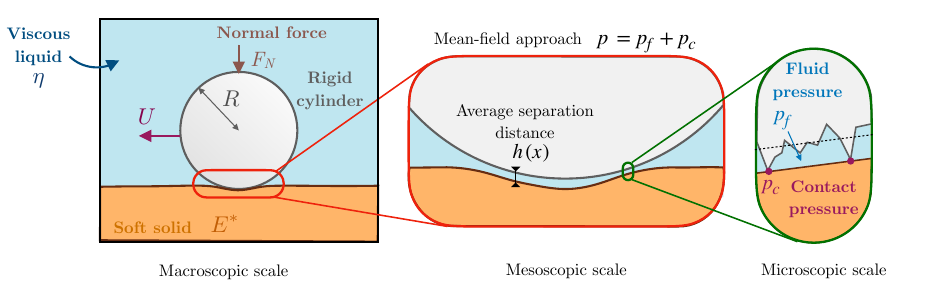}}	
\caption{\label{fig2} \textbf{Schematic of the homogenization method.} A rigid rough cylinder is sliding along a soft solid immersed in a viscous liquid. We model the contact area at the mesoscopic scale using a mean-field approach (see red zoomed region), where the asperities (see green zoomed region) are averaged spatially to a contact pressure. }
\end{figure}

The central difficulty in mixed-lubrication modeling is that it must simultaneously describe regions of fluid flow and regions of solid–solid contact. Classical Navier–Stokes–based models with no-slip boundary conditions, such as lubrication theory, cannot handle sliding contact: as the local film thickness tends to zero, the shear rate diverges. Within the Reynolds equation, this manifests as a singularity in the pressure field when the gap vanishes. This issue is well known in fluid dynamics and wetting, where it appears as the contact-line singularity~\cite{bonn2009wetting,snoeijer2013moving,kansal2024viscoelastic}. To circumvent this difficulty and capture both asperity contact and hydrodynamic pressure, homogenization mean-field approaches have been proposed (Fig.~\ref{fig2}), in which the interfacial pressure is split at a mesoscopic scale into a contact contribution and a hydrodynamic contribution~\cite{johnson1972simple,persson2009transition,persson2011lubricated,scaraggi2011lubrication}.

In this work, we analyze one of the most simple mean-field models of mixed lubrication, following the framework pioneered by Persson and Scaraggi~\cite{persson2009transition}. We choose to employ a minimal model with the aim of obtaining exact asymptotic expressions of the friction and providing rigorous criteria of lubrication transitions. The paper is organized as follows. Section~\ref{sec:model} introduces the model and governing dimensionless parameters. Section~\ref{sec:limiting_cases} addresses the limiting cases of boundary and hydrodynamic regimes. Section~\ref{sec:lubrication_transition} presents an asymptotic analysis of the frictional response around both the boundary lubrication and hydrodynamic lubrication regimes. These analyses provide a characterization of the transitions into the mixed regime. Conclusions and outlook are discussed in Section~\ref{sec:conclusion}.

\section{Mixed-lubrication mean-field model}
\label{sec:model}

We introduce a mean-field description of mixed lubrication that closely follows the framework proposed in Ref.~\cite{persson2009transition}. The model captures the coexistence of solid contact at asperity summits and hydrodynamic lubrication in the intervening fluid film. In section~\ref{subsec:formulation}, we first introduce the physical setting and the main modeling assumptions, before deriving the governing equations. A dimensional analysis is then carried out in Sec.~\ref{subsec:dimensional_analysis} to identify the relevant control parameters and limiting regimes.

\subsection{Formulation}
\label{subsec:formulation}

The model system consists of an infinitely long rigid cylinder of radius $R$ sliding at constant speed $U$ over a soft, planar elastic substrate (Fig.~\ref{fig2}(a)). The contact is fully immersed in a viscous liquid of viscosity $\eta$, and a prescribed normal load per unit length $F_N$ is applied to the cylinder.

Owing to surface roughness, the applied load is transmitted partly through direct solid contact at asperity summits and partly through the hydrodynamic pressure generated in the lubricating liquid film (Fig.~\ref{fig2}(c)). Rather than resolving the roughness explicitly, a spatially homogenized (mean-field) description is adopted at a mesoscopic scale (Fig.~\ref{fig2}(b)). Within this framework, the effects of roughness are accounted for through the steady-state, spatially averaged separation distance $h(x)$ and the corresponding averaged total pressure field $p(x)$.

The central assumption of the model is that, at any position $x$ along the contact, the total pressure can be decomposed linearly into a contact contribution $p_c(x)$ and a fluid contribution $p_f(x)$,
\begin{equation}
\label{eq:mean-field}
p(x) = p_c(x) + p_f(x),
\end{equation}
where the two pressure components are treated as independent and are both related to the average separation distance $h(x)$. 
This decomposition should be understood as a consequence of the homogenized description rather than as a local partition of the pressure at the microscopic scale. At the microscopic level, the interface consists of a heterogeneous distribution of asperity contacts and fluid-filled gaps, whose geometry continuously evolves during sliding. Rather than resolving this complex moving-boundary problem, the model replaces it with statistically averaged contact and fluid pressures, both expressed as functions of the mean interfacial separation $h(x)$. In this sense, the model constitutes a homogenization of the microscopic contact problem.

As a consequence, the model does not capture phenomena associated with the detailed morphology and dynamics of the contact patches, such as percolation, flow-channel connectivity, or contact-line motion. Accordingly, the present asymptotic analyses are expected to be most reliable in the boundary- and hydrodynamic-lubrication limits, whereas the transition criteria inherit the limitations of the underlying mean-field approximation when the fluid thickness becomes comparable to the roughness scale. A similar homogenization framework has been introduced in the continuum modeling of suspensions~\cite{jackson1997locally}.

The contact pressure is modeled using Persson’s contact mechanics theory for randomly rough, self-affine surfaces~\cite{persson2007relation}. Within this framework, the local contact pressure depends exponentially on the average separation distance between the surfaces,
\begin{equation}
\label{eq:contact_pressure}
p_c(x) = \beta E^* \exp\!\left(-\alpha \frac{h(x)}{h_{\mathrm{rms}}}\right),
\end{equation}
where $h_{\mathrm{rms}}$ is the root-mean-square roughness amplitude, $\alpha$ and $\beta$ are numerical coefficients of order unity determined by the full roughness spectrum, and $E^* = E/(1-\nu^2)$ is the effective elastic modulus, with $E$ and $\nu$ the Young’s modulus and Poisson ratio of the substrate. The exponential dependence is not empirical, but corresponds to the small-pressure asymptotic solution of Persson's contact mechanics theory. It is therefore expected to remain valid only while the mean separation is sufficiently larger than the roughness amplitude, whereas noticeable deviations occur as $h/h_\mathrm{rms} = O(1)$ or smaller, where the contact pressure eventually diverges as complete contact is approached~\cite{persson2007relation}.

Then, the fluid pressure is assumed to obey the Reynolds (thin-film) equation, widely employed for lubrication flows. In the present steady, one-dimensional configuration, this reduces to~\cite{snoeijer2013similarity}
\begin{equation}
\label{eq:thin-film}
p_f'(x) = 6\eta U\,\frac{h(x) - h^*}{h^3(x)},
\end{equation}
where $h^*$ is a constant proportional to the volumetric flux of fluid passing through the contact. We further neglect the effect of roughness on the fluid flux, often accounted for through so-called flow factors~\cite{patir1978,patir1979application}, and assume that the homogenized fluid pressure obeys the classical Reynolds (thin-film) equation.

The imposed normal load is carried by the total pressure, which imposes the integral constraints
\begin{equation}
\label{eq:load}
\int_{\mathbb{R}} p(x)\,\mathrm{d}x = F_N.
\end{equation}

The mean separation $h(x)$ results from the superposition of the undeformed geometry and the elastic deformation of the substrate induced by the total pressure field. Assuming plane-strain elasticity, the gap profile reads
\begin{equation}
\label{eq:elasticity}
h(x) = z_0 + \frac{x^2}{2R}
- \frac{2}{\pi E^*} \int_{\mathbb{R}} p(y)\,\ln|x-y|\,\mathrm{d}y,
\end{equation}
where $z_0$ denotes the vertical position of the cylinder apex and the second term $x^2/(2R)$ the local parabolic shape of the cylinder, while the last term represents the elastic deformations. In total, the equations~\eqref{eq:mean-field}–\eqref{eq:elasticity} involve four fields ($p$, $p_f$, $p_c$, and $h$) and two unknown constants, $h^*$ and $z_0$. These constants are determined by imposing the global force balance~\eqref{eq:load} together with the fluid pressure boundary conditions $p_f(\pm\infty)=0$, so that the system is well-posed. 

Once the separation and pressure fields are known, the tangential stress acting on the cylinder can be evaluated and integrated to obtain the total friction force. Normalizing this force by the applied normal load defines the friction coefficient $\mu$. As for the normal pressure, the tangential stress is decomposed into a solid (contact) contribution and a fluid (viscous) contribution. This decomposition directly translates into a corresponding decomposition of the friction coefficient,
\begin{equation}
\mu = \mu_c + \mu_f.
\end{equation}
The tangential stress transmitted through solid contact is assumed to obey a local Coulomb-like friction law and is therefore proportional to the local contact pressure. As a result, the contact contribution to the friction coefficient reads
\begin{equation}
\mu_c =  \frac{\mu_0}{F_N}\int_{\mathbb{R}} p_c(x)\,\mathrm{d}x,
\end{equation}
where $\mu_0$ denotes the solid friction coefficient associated with direct asperity contacts, i.e. the friction coefficient that would be measured if the applied load were entirely supported by solid-solid contact. Within the present model, $\mu_0$ is treated as an effective input parameter, assumed independent of the normal load and sliding velocity, although its value may in general depend on the roughness characteristics of the interface.
Since $\mu_0$ enters the problem only through the friction law, and not explicitly in Eqs.\eqref{eq:mean-field}-\eqref{eq:elasticity}, it does not affect the pressure or separation distance spatial distributions. Changing its value therefore modifies the predicted friction coefficient, but leaves the underlying contact and lubrication fields unchanged.

The fluid contribution arises from the tangential traction exerted by the fluid on the cylinder, which contains both the viscous shear stress and the tangential projection of the fluid pressure. Within the lubrication approximation, the resulting friction coefficient is
\begin{equation}
\mu_f = \frac{1}{F_N}\int_{\mathbb R} \left(-p_f\frac{x}{R}-\eta\left.\frac{\partial v}{\partial z}\right|_{z=z_c} \right)\, \mathrm{d}x.
\end{equation}
where $z_c = z_0 + x^2/(2R)$ denotes the cylinder surface and $v(x,z)$ the fluid velocity field. Using standard lubrication theory, this expression simplifies to
\begin{equation}
\label{eq:friction}
\mu_f =\frac{\eta U}{F_N} \int_{\mathbb R} \frac{1}{h}\left[ 1-3\frac{h-h^*}{h}+6\frac{h-h^*}{h^2}\frac{x^2}{2R}\right] \mathrm{d}x.
\end{equation}
which will be used in the following to construct the Stribeck curves predicted by the present mean-field model. Compared with the expression used in Ref.~\cite{persson2009transition}, the present form additionally accounts for the tangential projection of the hydrodynamic pressure on the curved cylinder surface.

\subsection{Dimensional analysis and phase diagram}
\label{subsec:dimensional_analysis}
\begin{figure}[h!]
\centerline{\includegraphics{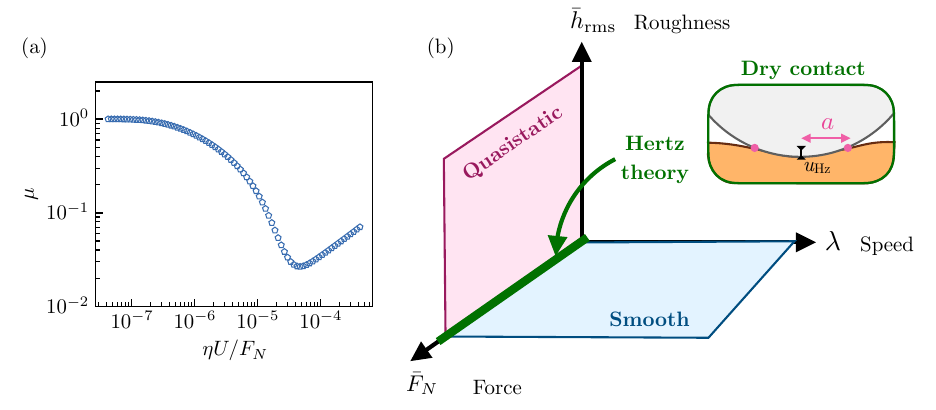}}	
\caption{\label{fig3} \textbf{Numerical Stribeck curve and phase diagram.} (a) Friction coefficient versus the Hersey number for $\bar{h}_\mathrm{rms} = 10^{-4}$, $\bar{F}_N = 0.1$ and $\mu_0 = 1$.  (b) Schematic phase diagram of the lubrication problem in the three-dimensional parameter space defined by the dimensionless speed $\lambda$, load $\bar{F}_N$ and roughness $h_\mathrm{rms}$. The diagram highlights key asymptotic regimes: the smooth limit ($\bar{h}_\mathrm{rms}=0$, blue plane), corresponding to classical elastohydrodynamic lubrication; the quasistatic limit ($\lambda = 0$, pink plane), where the friction is purely contact-dominated; and their intersection (green line), representing the Hertzian indentation problem for dry contacts, as schemed in the green box. }
\end{figure}
There are multiple length and pressure scales in the problem, arising from elasticity, lubrication forces, surface roughness, and the imposed normal load. The purpose of this subsection is to identify the relevant dimensionless parameters governing the lubrication transition.

The system is characterized by six physical parameters: the sliding velocity $U$, normal force $F_N$, surface roughness amplitude $h_\mathrm{rms}$, cylinder radius $R$, effective elastic modulus $E^*$, and fluid viscosity $\eta$. Dimensional analysis (Buckingham $\Pi$ theorem) therefore implies the existence of three independent dimensionless groups. We choose to express these groups as a dimensionless speed, force, and roughness amplitude, as these correspond directly to experimentally controlled parameters.

Several nondimensionalization choices are possible. Following Snoeijer et al.~\cite{snoeijer2013similarity}, we adopt the elastic scales provided by Hertz theory for the dry contact between a smooth rigid cylinder and an elastic half-space. Specifically, we use as characteristic scales the maximum Hertzian pressure $p_{\mathrm{Hz}}$, the typical elastic deformation $u_{\mathrm{Hz}}$, and the half-width of the dry contact $a$, which are given by~\cite{johnson1987contact}
\begin{equation}
\label{eq:Hertz_scales}
p_\mathrm{Hz} = \frac{2F_N}{\pi a} = \left(\frac{F_N E^*}{\pi R}\right)^{1/2},\quad \quad a = \left(\frac{4F_N R}{\pi E^*}\right)^{1/2}, \quad \quad u_\mathrm{Hz} = \frac{a^2}{R} = \frac{4F_N}{\pi E^*}.
\end{equation}
Using these Hertzian scales, we first define a dimensionless speed quantifying the relative importance of lubrication and elastic forces. Substituting the above scales into the thin-film equation~\eqref{eq:thin-film} yields
\begin{equation}
\lambda = \frac{6\eta U a}{u_\mathrm{Hz}^2 p_\mathrm{Hz}} = \frac{3\pi^2}{4}\frac{\eta U R E^*}{F_N^2},
\end{equation}
which is proportional to the classical Hersey number, up to the factor $3\pi^2 R E^/(4F_N)$ arising from the elastic normalization.

Substituting the same scales into the contact pressure relation~\eqref{eq:contact_pressure} yields two additional dimensionless parameters,
\begin{equation}
\bar{F}_N = \frac{1}{\pi \beta^2}\frac{F_N}{R E^*}, \qquad
\bar{h}_\mathrm{rms} = \frac{1}{4\alpha\beta^2}\frac{h_\mathrm{rms}}{R},
\end{equation}
which correspond respectively to the applied normal force normalized by the elastic stiffness and to the roughness amplitude normalized by the cylinder radius. 
Alternatively, $\bar{F}_N \propto  (p_\mathrm{Hz}/(\beta E^*))^2$, so that $\bar{F}_N$ compares the Hertzian contact pressure to the characteristic pressure scale appearing in Persson's contact law.
The numerical prefactors $\alpha$ and $\beta$ in \eqref{eq:contact_pressure} have been absorbed into the definition of the dimensionless groups. 

Appendix~\ref{app:dimensionless} presents the dimensionless set of equations obtained once normalizing by the aforementioned scales. This set of equations is then solved numerically using a finite-difference scheme. For a given solution at prescribed values of $(\lambda,\bar{F}_N,\bar{h}_\mathrm{rms})$, a custom continuation algorithm is employed to systematically explore the parameter space. To illustrate the ability of the present mean-field model to capture the lubrication transition, Fig.~\ref{fig3}(a) shows a representative Stribeck curve obtained for $\bar{h}_\mathrm{rms}=10^{-4}$, $\bar{F}_N=0.1$, and $\mu_0=1$. The classical Stribeck behavior is recovered, including (i) a boundary-lubrication regime at vanishing sliding velocity with an approximately constant friction coefficient for $\eta U/F_N < 10^{-7}$, (ii) a mixed-lubrication regime in which the friction coefficient decreases with increasing speed for $10^{-7} <\eta U/F_N < 3 \times 10^{-5}$, and (iii) a hydrodynamic-lubrication regime at sufficiently large speeds $\eta U/F_N > 3 \times 10^{-5}$, where the friction coefficient increases with both sliding speed and fluid viscosity.

Altogether, we obtain a three-dimensional phase space that is drawn schematically in Fig.~\ref{fig3}, where the three axes correspond to the dimensionless speed, normal force and roughness. On the phase diagram, we can already identify some specific regions corresponding to well defined regimes of friction. 
\begin{itemize}
\item The first limit of interest is the regime of validity of the Hertz-contact solution, which is the case of smooth surface $\bar{h}_\mathrm{rms} = 0$ and zero sliding speed $\lambda = 0$. On the phase diagram of Fig.~\ref{fig3}, this corresponds to the axis highlighted with a green line. We briefly recall the pressure and separation distance profiles of the Hertz theory in Sec.~\ref{subsec:Hertz} as it will arise later in the asymptotic analysis.

\item The second one is the smooth limit, meaning the blue plane $\bar{h}_\mathrm{rms} = 0$ in Fig.~\ref{fig3}. In that case, no direct contact is allowed and the friction is only characterized by the hydrodynamic pressure, which corresponds to the classical elastohydrodynamic lubrication regime~\cite{dowson2014elasto}. In Sec.~\ref{subsec:smooth}, we provide a derivation of the hydrodynamic friction force in this case, that is built on known asymptotic solutions of the problem. 

\item The last one is the quasistatic limit, meaning pink plane $U = \lambda = 0$. In this situation, our model reduces to the indentation of a rough cylinder on a soft surface. This corresponds to the boundary lubrication regime, where the friction is given by the Coulomb friction coefficient. In Sec.~\ref{subsec:quasistatic}, we characterize the separation distance profile in that regime, as it will be useful later to describe the transition between regimes. 

\end{itemize}

To sum up, in this section, we have identified the three governing dimensionless numbers of our model, which can be seen as dimensionless speed, normal force and roughness. Furthermore, we introduce the two limiting cases of smooth and quasistatic, which correspond to the elastohydrodynamic and boundary lubrication regimes respectively. In the next section, we will focus on these two limits of the model.

\section{Limiting cases}
\label{sec:limiting_cases}
Before investigating the mixed-lubrication regime in the next section, we first focus here on the boundary and hydrodynamic lubrication regime. Those arise naturally in the limiting cases that have been identified in Fig.~\ref{fig3}(b). More specifically, we will describe here the Hertzian contact, smooth limit (hydrodynamic) and quasistatic limit (boundary) in this order. 

\subsection{Hertz solution $U, h_\mathrm{rms}\to 0$}
\label{subsec:Hertz}
The first and simplest limiting case is the famous Hertz solution of contact mechanics, meaning the static indentation of a smooth cylinder on a soft surface. The Hertz solution is expected to emerge both in the quasistatic ($U \to 0$) and the smooth limit ($h_\mathrm{rms} \to 0$). The latter can be found in textbooks for the pressure distribution and separation distance profile for the pressure and thickness profile that reads~\cite{johnson1987contact}
\begin{equation}
\label{eq:Hertz}
p(x) = p_\mathrm{Hz}\left(1-\frac{x^2}{a^2}\right)^{1/2}\Theta\left(a^2-x^2\right), \quad  h(x) =  \frac{u_\mathrm{Hz}}{2}\left(\left\lvert \frac{x}{a} \right\rvert  \sqrt{\left(\frac{x}{a}\right)^2-1} -\ln\left[ \left\lvert  \frac{x}{a} \right\rvert +\sqrt{\left(\frac{x}{a}\right)^2-1} \right]\right) \Theta\left(x^2-a^2\right),
\end{equation}
where we introduce the Heaviside function $\Theta(x>0) = 1$ and $\Theta(x<0) = 0$. We notice that the vertical position of the cylinder is equal to 
\begin{equation}
\label{eq:vertical_position_Hz}
z_0 = -u_\mathrm{Hz} \left(1 + \ln 4\right)/4 \approx -0.59657 \, u_\mathrm{Hz},
\end{equation}
which can be found evaluating Eq.~\eqref{eq:elasticity} at $x=0$. 

\subsection{Smooth limit ($h_\mathrm{rms}\to 0$): hydrodynamic lubrication}
\label{subsec:smooth}
\begin{figure}[h!]
\centerline{\includegraphics{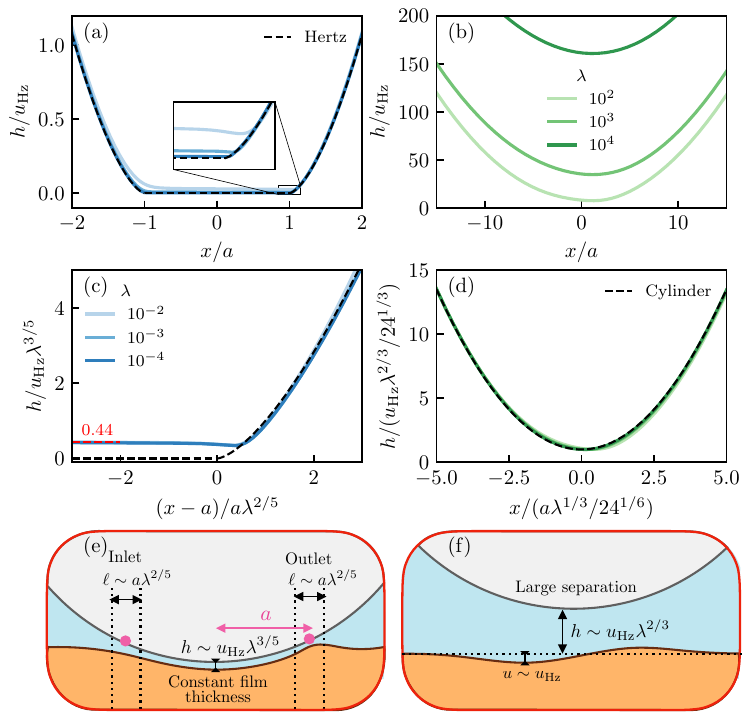}}	
\caption{\label{fig4} \textbf{Separation distance profiles in the smooth limit.} Dimensionless separation distance as a function of the lateral position for various values of $\lambda$ (indicated with colors) both in the $\lambda \ll 1$ (a) and $\lambda \gg 1$ (b) limits. The inset in (a) displays a zoom in the outlet region, highlighting the $\lambda$ dependence of the separation distance in the central region. In the panels (c) and (d), we plot the same data as in (a)-(b) but with the appropriate $\lambda$ rescaling to highlight the asymptotic solutions. The schematics in (e) and (f) highlight the structure of the lubrication film. The pink dot shows the location of the Hertz contact radius.}
\end{figure}

In the limit of a smooth cylinder, formally corresponding to a zero root-mean-square roughness in the present model, the lubrication pressure entirely accommodates the normal load while the contact pressure is null. In this situation, the model of Sec.~\ref{subsec:formulation} reduces to the elastohydrodynamic lubrication model~\cite{venner2000multi,lugt2011review,greenwood2020elastohydrodynamic}, governed by a single dimensionless number that is $\lambda$ here. We notice that some other papers used different renormalization scales, but an equivalent dimensionless number $M = F_N/(\eta U E^* R)^{1/2} \propto \lambda^{-1/2}$~\cite{moes1992optimum}. 
We stress that we restrict ourselves to the isoviscous case where the viscosity is pressure independent, and we do not consider any possible cavitation at negative pressure~\cite{greenwood2020elastohydrodynamic}. 

This case has been widely discussed in the literature, and in particular some asymptotic solutions have been proposed in the $\lambda \ll 1$ and $\lambda \gg 1$ limits, as we will discuss below. We illustrate those asymptotic solutions in the Fig.~\ref{fig4} showing the film thickness profiles for various $\lambda$, where the $\lambda \ll 1$ (resp. $\lambda \gg 1$) limit is shown on the left (resp. right), which correspond to large-force/small-speed and large-speed/low-force respectively. Alternatively, Essink et al. introduced the terminology large ($\lambda \ll 1$) and small ($\lambda \gg 1$) deformation regime~\cite{essink2021regimes}, that we are going to employ in the following subsections. 

\subsubsection{Large deformation limit ($\lambda \ll 1$)}
\label{subsubsec:large_deformation_smooth}

Clearly, in the $\lambda \to 0$ limit, the separation distance profile converges toward the Hertz solution Eq.~\ref{eq:Hertz} (see Fig.~\ref{fig4}(a)). In this regime, the film thickness is much smaller than the typical elastic displacement $u_\mathrm{Hz}$, which justifies the terminology \textit{large deformation limit}. Nevertheless, the Hertz solution alone is not sufficient to find the friction force as it does not predict the film thickness within the contact region. The inset in Fig.~\ref{fig4}(a) represents a zoom of the outlet region that clearly illustrates that the film thickness remains $\lambda$ dependent. 

\begin{figure}[h!]
\centerline{\includegraphics{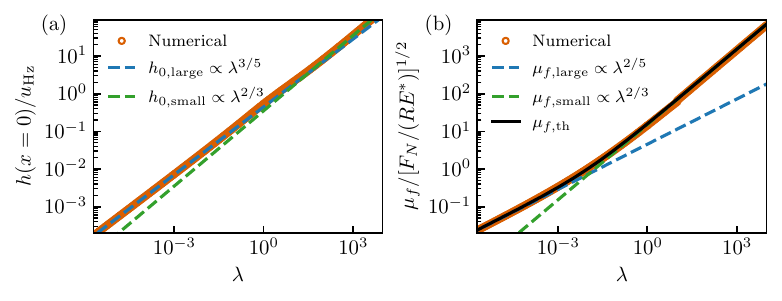}}	
\caption{\label{fig5} \textbf{Central film thickness and hydrodynamic lubrication friction in the smooth limit.} (a) Dimensionless central film thickness $h(x=0)/u_\mathrm{Hz}$ versus $\lambda$, where the blue and green dashed lines show \eqref{eq:central_thickness_large} and \eqref{eq:film_thickness_small_deformation} respectively. (b) Dimensionless hydrodynamic friction coefficient versus the dimensionless speed $\lambda$ in logarithmic scales, where the circles show the numerical integration of \eqref{eq:hydrofriction_full} and the solid lines the interpolation function \eqref{eq:interpolated_hydrodynamic_friction}. The blue and green dashed lines display the asymptotic law \eqref{eq:hydrofriction_large} and \eqref{eq:hydrofriction_small} respectively. }
\end{figure}

Using asymptotic matching methods, Snoeijer et al.~\cite{snoeijer2013similarity} characterized the structure of the film thickness in this limit, as schematically illustrated in Fig.~\ref{fig4}(e). In the central region ($\lvert x \rvert < a$), the film thickness is nearly uniform and given by
\begin{equation}
\label{eq:central_thickness_large}
h(\lvert x \rvert < a) = h_{0,\rm large} \approx 0.44 u_\mathrm{Hz} \lambda^{3/5}.
\end{equation}
Figure~\ref{fig5}(a) shows the central film thickness $h(x=0)$ as a function of $\lambda$, which is well described by the asymptotic law~\eqref{eq:central_thickness_large}.

Near the edges of the central region, at $x=\pm a$, two universal self-similar solutions develop, each with a lateral extent scaling as $a\lambda^{2/5}$. These boundary-layer–like regions smoothly connect the central flat film to the Hertzian profile outside the contact area ($\lvert x \rvert > a$). Although the inlet ($x=-a$) and outlet ($x=a$) solutions differ, they share identical characteristic scales. To illustrate the universal nature of the outlet solution, we replot in Fig.~\ref{fig4}(c) the separation distance profiles from Fig.~\ref{fig4}(a), where the thickness is rescaled by the central film scaling and the lateral scale by the boundary-layer thickness. This rescaling yields an almost perfect collapse of the numerical data, confirming the existence of a universal self-similar solution.

To determine the hydrodynamic friction force (see Eq.~\eqref{eq:friction}), it remains to compute the integration constant $h^*$, which physically corresponds to the dimensionless liquid flux. Integrating Eq.~\eqref{eq:thin-film} over the entire domain and applying the pressure boundary conditions $p(\pm\infty)=0$ yields the closed-form expression~\cite{yariv2024hydrodynamic}
\begin{equation}
\label{eq:hstar}
h^* = \frac{\displaystyle \int_{\mathbb{R}} h^{-2}(x),\mathrm{d}x}{\displaystyle \int_{\mathbb{R}} h^{-3}(x),\mathrm{d}x}.
\end{equation}
Given the spatial structure of the film thickness, both integrals in Eq.~\eqref{eq:hstar} are dominated by the contribution from the central flat region. We may therefore approximate
\begin{equation}
\int_{\mathbb{R}}h^{-n}(x) \, \mathrm{d}x \approx \int_{-a}^ah^{-n}(x) \, \mathrm{d}x\approx 2 a h_{0, \rm large}^{-n} ,\quad \text{for}\quad  n = 1, 2, 3.
\end{equation}
which immediately yields $h^* = h_{0,\rm large}$.

The hydrodynamic friction coefficient then follows from integrating the viscous shear stress in Eq.~\eqref{eq:friction}, leading to
\begin{equation}
\label{eq:hydrofriction_full}
%\mu_f = \frac{\eta U}{F_N}\int_{\mathbb{R}} \frac{1}{h(x)}\left[1 + 3\left(1 - \frac{h^}{h(x)}\right)\right]\mathrm{d}x.
\mu_{f} = \frac{\eta U}{F_N}\int_\mathbb{R}  \frac{1}{h(x)} \left[ 1-3\frac{h-h^*}{h}+6\frac{h-h^*}{h^2}\frac{x^2}{2R}\right] \mathrm{d}x.
\end{equation}
We now derive an asymptotic expression for the fluid friction coefficient in the large-deformation limit, denoted $\mu_{f,\rm large}$. As for the flux calculation, the integral in Eq.~\eqref{eq:hydrofriction_full} is dominated by the central region, where the film thickness is uniform. Since $h^* = h_{0,\rm large}$, the term in brackets vanishes in this region. Combining these approximations, we obtain
\begin{equation}
\label{eq:hydrofriction_large}
\mu_{f,\rm large} \simeq \frac{\eta U}{F_N}\frac{2a}{0.44u_\mathrm{Hz}\lambda^{3/5}}
\propto
\left(\frac{F_N R}{E^*}\right)^{1/2}\lambda^{2/5}.
\end{equation}
The numerical evaluation of the full hydrodynamic friction from Eq.~\eqref{eq:hydrofriction_full}, shown in Fig.~\ref{fig5}(b) (orange circles), is in excellent agreement with the asymptotic prediction~\eqref{eq:hydrofriction_large} (dashed blue line) in the $\lambda \to 0$ limit.

\subsubsection{Small deformation limit ($\lambda \gg 1$)}
\label{subsubsec:small_deformation_smooth}
The film thickness profiles have very different shapes qualitatively in the large $\lambda$ limit (see Fig.~\ref{fig4}(b)). Oppositely to the $\lambda \ll 1$, the film thickness is now much larger than the typical elastic deformations. Skotheim and Mahadevan have solved the elastohydrodynamic equations by performing a regular perturbation expansion~\cite{skotheim2005soft}. The elastic deformations can be neglected  with respect to the minimum film thickness $h_{0, \rm small}$ to leading order and such that the film thickness and pressure field are well approximated by 
\begin{equation}
\label{eq:fields_small_deformation}
h(x) \approx h_{0, \rm small}  + \frac{x^2}{2R}, \quad \quad p(x) \approx -2\eta U \frac{x}{(h_{0, \rm small} +x^2/2R)^2}, \quad \quad h^* = \frac{4 h_{0, \rm small} }{3}.
\end{equation}
The resolution of the next leading order equation together with the integral constraint on the pressure field provides the missing condition to get $h_{0, \rm small} $ that gives 
\begin{equation}
\label{eq:film_thickness_small_deformation}
h_{0, \rm small} = \left(\frac{3\pi\eta^2U^2R^2}{2E^*F_N}\right)^{1/3} = u_\mathrm{Hz} \frac{\lambda^{2/3}}{24^{1/3}},
\end{equation}
where we recover that $h_{0, \rm small}  \gg u_\mathrm{Hz}$ when $\lambda$ is large. In Fig.~\ref{fig4}(d), we show the same separation distance profiles as in Fig.~\ref{fig4}(b), but where the film thickness is rescaled by $h_{0, \rm small} $ while the lateral scale is rescaled by $\sqrt{Rh_{0, \rm small} }$, which leads a perfect collapse, where the black dashed lines shows a parabolic law $y = 1 + x^2/2$ corresponding to the rigid cylinder interface. Additionally, the asymptotic expression~\eqref{eq:film_thickness_small_deformation} perfectly describes the numerical central film thickness at large $\lambda$, as shown in Fig.~\ref{fig5}(a) with the green dashed line.

Now that we have determined the morphology of the separation distance profile, we can obtain an approximate expression for the hydrodynamic friction in the $\lambda \to \infty$ limit. We inject \eqref{eq:fields_small_deformation} and \eqref{eq:film_thickness_small_deformation} into \eqref{eq:friction} and perform the integration to get
\begin{equation}
\label{eq:hydrofriction_small}
\mu_{f,\rm small} = \frac{2 \sqrt{2}\pi \eta U R^{1/2}}{F_N h_0^{1/2}} \propto \left(\frac{F_N R}{E^*}\right)^{1/2} \lambda^{2/3}, 
\end{equation}
where a different speed dependence is found as compared to the large deformation regime \eqref{eq:hydrofriction_large}. Again, the asymptotic expression~\eqref{eq:hydrofriction_small} (green dashed lines in Fig.~\ref{fig5}(b)) describes very well the full numerical integration of the hydrodynamic friction in the $\lambda \to \infty$ limit.

\subsubsection{General model of the hydrodynamic friction}
\label{subsubsec:general}
In summary, the hydrodynamic friction follows two distinct asymptotic expressions with the dimensionless speed for $\lambda \to (0, \infty)$, each of them with a non-trivial scaling exponent. We wish to have a general prediction for $\mu_{f,\rm th}$ over the entire range of $\lambda$. We propose here an empirical function that interpolates the two asymptotic expressions as
\begin{equation}
\label{eq:interpolated_hydrodynamic_friction}
\mu_{f,\rm th} = \sqrt{\mu_{f,\rm small}^2 +\mu_{f,\rm large}^2}.
\end{equation}
The interpolation function \eqref{eq:interpolated_hydrodynamic_friction} is plotted with a solid black line and provides an excellent description of the hydrodynamic friction, including in the transition region. 

\subsection{Quasistatic limit ($U\to 0$): boundary lubrication}
\label{subsec:quasistatic}
\begin{figure}[h!]
\centerline{\includegraphics{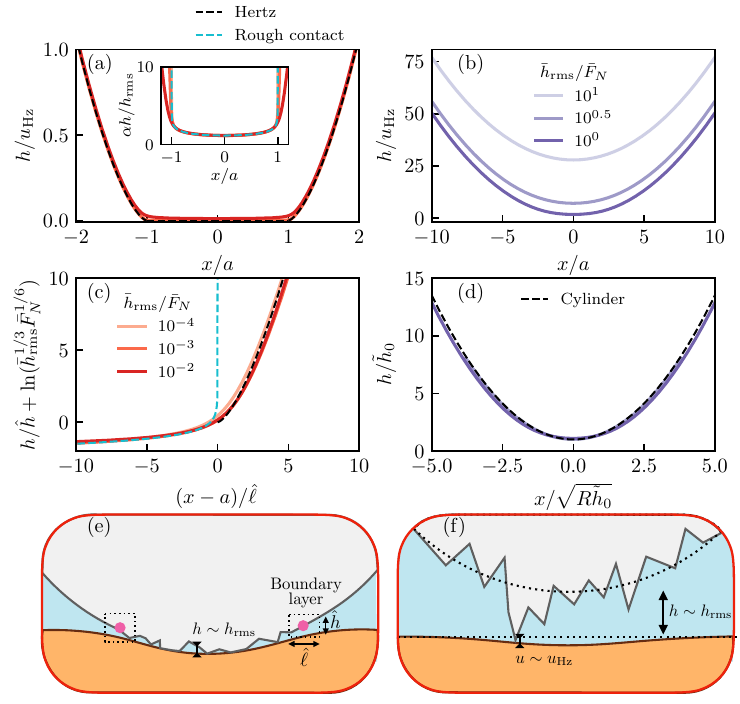}}	
\caption{\label{fig6} \textbf{Separation distance profiles in the quasistatic limit.} Rescaled separation distance profiles for large indentation (a) and small indentation (b). The different colors code for the dimensionless roughness parameter. The normalized load has been fixed to $\bar{F}_N = 0.1$. The inset in (a) shows a zoom in the contact region, where the separation are rescaled by the roughness. The cyan dashed lines shows the rough contact expression~\eqref{eq:central_region_contact} while the black dashed lines show the Hertz solution~\eqref{eq:Hertz}. The panel (c) show a zoom in the outlet region where the data are rescaled with the self-similar scales (see Appendix~\ref{app:boundarylayer}). In the panel (d), the data of (b) are rescaled vertically by $\tilde{h}_0$ (see Eq.~\eqref{eq:separation_small_indentation}) and horizontally by $\sqrt{a\tilde{h}_0}$. The schematics in (e) and (f) highlight the structure of the separation distance. We notice that we wrote $h \sim h_\mathrm{rms}$ to simplify, although logarithmic corrections in $\bar{F}_N$ and $\bar{h}_\mathrm{rms}$ exists for the typical separation distance scale.}
\end{figure}

We now move to the boundary lubrication limit, for zero sliding speed, meaning $\lambda = 0$, hence $p_f = 0$. In this situation, the hydrodynamic pressure is zero and the total friction is directly given by the Coulomb friction, i.e. $\mu = \mu_0$. We wish here to discuss the shape of the averaged separation and its dependence on the two dimensionless numbers $\bar{F}_N$ and $\bar{h}_\mathrm{rms}$. The main physical parameter involves the comparison between the typical Hertzian elastic indentation to the size of the roughness, which can be quantify as the ratio $\bar{h}_\mathrm{rms}/\bar{F}_N = h_\mathrm{rms}/(\alpha u_\mathrm{Hz})$. We plot in Fig.~\ref{fig6}(a)-(b) the pressure profiles for various values of $\bar{h}_\mathrm{rms}/\bar{F}_N$ and we recover a very similar phenomenology as for the smooth limit of section~\ref{subsec:smooth}. The $\bar{h}_\mathrm{rms}/\bar{F}_N \ll 1$ limit corresponds to large indentation with respect to the roughness scale and is displayed on the left while the small indentation limit is shown on the right. In the next couple of subsections, we will provide a quantitative description of the average separation distance profile in both of these limiting cases.

\subsubsection{Large indentation limit ($\bar{h}_\mathrm{rms}/\bar{F}_N \ll 1$)}
\label{subsubsec:large_quasi}
The first and most important limit in real applications is the situation where asperities are much smaller than the typical indentation scale, i.e. $\bar{h}_\mathrm{rms}/\bar{F}_N \ll 1$. In this case, the separation distance profile converges towards the Hertz profile, as shown in Fig.~\ref{fig6}(a). The inset in Fig.~\ref{fig6}(a) shows the same data rescaled by the root-mean-square roughness, and demonstrates that the separation distance in the central region scales with the typical roughness scale, up to a $\alpha$ factor. Indeed, the separation profile can be found analytically by injecting the Hertz pressure \eqref{eq:Hertz} into the contact equation \eqref{eq:contact_pressure}, leading to 
\begin{equation}
\label{eq:central_region_contact}
h(x) = -\frac{h_\mathrm{rms}}{2\alpha} \ln\left(\frac{F_N}{\beta^2 \pi R E^*} \left(1-\frac{x^2}{a^2}\right)\right).
\end{equation}
The solution \eqref{eq:central_region_contact} is shown in the inset of Fig.~\ref{fig6}(a) with blue dashed lines and labeled ‘‘rough contact'' and perfectly describes the numerical data in the contact region $\lvert x \rvert < a$. 

We wish to make a couple of remarks at this point. First, we set arbitrarily the dimensionless normal force to a value of $\bar{F}_N = 0.1$ in the data shown in Fig.~\ref{fig6}, which leads to a positive separation distance at the apex of the cylinder ($h/(h_\mathrm{rms}/\alpha) \approx \ln(10)/2 \approx 1.15$. Nevertheless, whenever $\bar{F}_N>1$, the expression~\eqref{eq:central_region_contact} leads to negative central separation distance as $h(x=0) \propto - \ln\left(\bar{F}_N\right)$, which is nonphysical. 
This behavior does not reflect a physical instability, but rather the breakdown of the exponential pressure-separation relation outside its range of validity. Being a large separation distance asymptotic expression, Eq.~\eqref{eq:contact_pressure} predicts a finite contact pressure at zero separation; its inverse therefore necessarily yields negative separations once this finite pressure is exceeded. A more realistic contact law would instead become increasingly stiff as complete contact is approached, preventing such unphysical solutions. Furthermore, values of $\bar{F}_N = O(1)$ also correspond to large elastic deformations, where the small-deformation approximation and the parabolic description of the cylinder are expected to lose validity. The condition $\bar{F}_N <1$ should therefore be viewed as a qualitative indication of the regime of validity of the present model rather than as a sharp physical threshold.

The second remark that we point out here is that the rough contact solution has a logarithmic singularity at the edge of contact $h(\lvert x \rvert \to a^-) \propto \ln(a-\lvert x \rvert)$. Hence, the rough contact solution must be regularized by a boundary layer at the edge of the contact zone $x\approx \pm a$, which interpolate between the rough contact solution~\eqref{eq:central_region_contact} and the Hertzian solution outside the contact region~\eqref{eq:Hertz}. This is analogous to the inlet/outlet solutions discussed for the elastohydrodynamic case of section \ref{subsubsec:large_deformation_smooth} and shown in Fig.~\ref{fig4}(c). In Appendix~\ref{app:boundarylayer}, we find the typical scales of this boundary layer, as well as the self-similar solutions.

We rescale the numerical profiles in Fig.~6(c), using those boundary-layer scales $\hat{h} = h_\mathrm{rms}/\alpha$, $\hat{\ell} = a (h_\mathrm{rms}/\alpha u_\mathrm{Hz})^{2/3}$ and $\hat{p} = p_\mathrm{Hz} (h_\mathrm{rms}/\alpha u_\mathrm{Hz})^{1/3}$ (see Appendix~\ref{app:boundarylayer}). The resulting collapse onto a single master curve confirms the existence of a universal self-similar boundary-layer solution that regularizes the logarithmic divergence and connects smoothly to the Hertzian outer profile.

\subsubsection{Small indentation limit ($\bar{h}_\mathrm{rms}/\bar{F}_N \gg 1$)}
\label{subsubsec:small_quasi}

The other limit $\bar{h}_\mathrm{rms}/\bar{F}_N \gg 1$ corresponds to the case where the typical Hertzian elastic indentation is much smaller than the typical roughness. In that case, the contact is governed by a few asperities as shown in the schematic of Fig.~\ref{fig6}(f). We show some separation profiles in Fig.~\ref{fig6}(b) in this limit. We recover the same phenomenology as in the Sec.~\ref{subsubsec:small_deformation_smooth}, where the separation distance largely exceeds the Hertz indentation scales as well as the typical elastic displacements. Therefore, we make the same assumption as in Sec.~\ref{subsubsec:small_deformation_smooth} and suppose that we can neglect elastic deformations in the separation distance to leading order and assume a parabolic separation distance profile as $h = \tilde{h}_0+x^2/(2R)$, where $\tilde{h}_0$ is an unknown distance. Injecting this Ansatz in the contact equation \eqref{eq:contact_pressure} allows to obtain the pressure profile $p(x) = -\beta E^* \exp\left[-\alpha(\tilde h_0 + x^2/(2R))/h_\mathrm{rms}\right]$ that is Gaussian. Lastly, we perform the integral of the resulting pressure profile over the entire domain which must equate normal load (see Eq. \eqref{eq:load}). This provide a condition that allows to solve for the unknown distance $\tilde{h}_0$ that is found to be 
\begin{equation}
\label{eq:separation_small_indentation}
\tilde{h}_0 = -\frac{h_\mathrm{rms}}{2\alpha}\ln \left(\frac{\pi \bar{F}_N^2}{8\bar{h}_\mathrm{rms}}\right) = - \frac{h_\mathrm{rms}}{2\alpha} \ln \left(\frac{F_N^2 \alpha}{2\pi \beta^2 Rh_\mathrm{rms} E^{*2}}\right).
\end{equation}
As expected, we find that the typical scale of the separation distance is the surface roughness. What is less obvious is the logarithmic corrections, both in dimensionless normal load and dimensionless roughness. In Fig.~\ref{fig6}(d), we plot the same separation distance profiles as in Fig.~\ref{fig6}(b) but rescaled by $\tilde{h}_0$ while the x-axis is rescaled by $\sqrt{\tilde{h}_0R}$. Again, we find a perfect collapse of all the numerical data, that confirms our analysis. 

To sum up, we discuss the shape of the average separation distance in the quasistatic limit of our contact model. The main controlled parameter is the ratio between the roughness to the Hertz indentation $\bar{h}_\mathrm{rms}/\bar{F}_N$. We found two asymptotic regimes, either at large or small indentation, respectively $\bar{h}_\mathrm{rms}/\bar{F}_N \to (0, \infty)$. In both cases, the typical scale of the separation distance is given by the roughness, with non-trivial logarithmic correction in the dimensionless normal load, as well as roughness. In what follows, we will focus on the limit of nearly smooth surface with $\bar{h}_\mathrm{rms}/\bar{F}_N < 10^{-2}$ and below. Nonetheless, it could be of interest in future work to extend the analysis to the opposite rough limit, and to surfaces with ordered roughness with a discrete spectrum. %where the contact pressure model could be adapted. %\textbf{[TO DO: some bibliography to add citation and be more precise.]}

\section{Lubrication transition}
\label{sec:lubrication_transition}
\begin{figure}[h!]
\centerline{\includegraphics{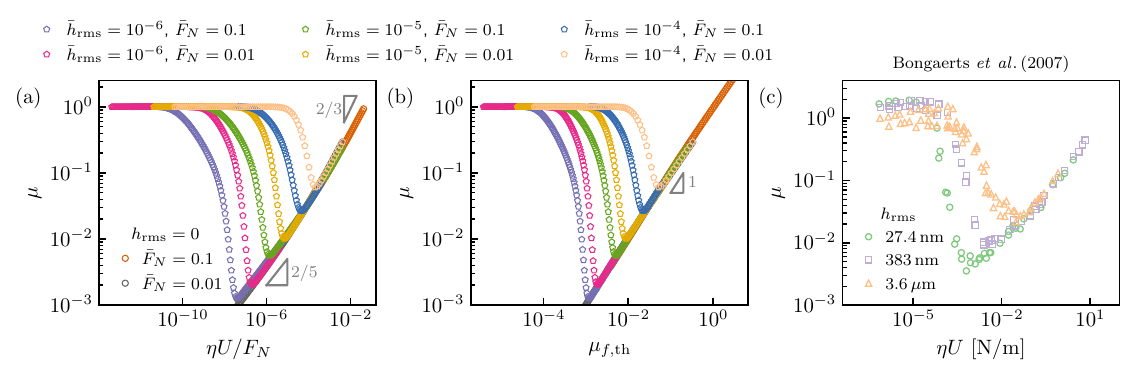}}	
\caption{\label{fig7} \textbf{Stribeck curve.} (a) Friction coefficient as function of the Hersey number in logarithmic scales, where the different colors correspond to varying contact parameters ($\bar{h}_\mathrm{rms}$ and $\bar{F}_N$) indicated in the legends. In the panel (b), the same data are plotted versus the prediction~\eqref{eq:interpolated_hydrodynamic_friction} of the hydrodynamic friction. The gray triangles indicate power-law scalings. (c) Experimental Stribeck curve reported in Ref.~\cite{bongaerts2007soft}. }
\end{figure}

\subsection{Stribeck curve}

We now turn to the full system of equations \eqref{eq:mean-field}–\eqref{eq:elasticity} to analyze the mixed-lubrication regime and the transitions between frictional states. Numerical simulations were performed by fixing the roughness and normal force while varying the dimensionless sliding speed over a large range (up to 10 orders of magnitude). Throughout, we set the Coulomb friction parameter to $\mu_0 = 1$ in all Stribeck curves. Changing $\mu_0$ modifies the overall curve shape but not the separation distance profiles, since $\mu_0$ does not explicitly appear in \eqref{eq:mean-field}–\eqref{eq:elasticity}.

Figure~\ref{fig7}(a) shows the Stribeck curve, i.e. the total friction coefficient as a function of the Hersey number $\eta U/F_N$. Regardless of the contact parameters ($\bar{h}_\mathrm{rms}$ and $\bar{F}_N$), the friction always transitions from a contact-dominated regime (boundary lubrication) at low speeds to a fluid-dominated regime (hydrodynamic lubrication) at high speeds, with a mixed regime of velocity weakening in between. Smoother surfaces shift the transition to lower speeds and reduce the overall friction, as expected. The same trend occurs for larger dimensionless normal forces. However, the precise transition speeds depend sensitively on both $\bar{h}_\mathrm{rms}$ and $\bar{F}_N$, and understanding these dependencies is a central goal of this section.

When plotted against the Hersey number, the hydrodynamic regime exhibits a tiny spread among the numerical curves, reminiscent of the fact that $\eta U/F_N$ is not the sole governing parameter. Indeed, the simulations recover two distinct asymptotic scalings $U^{2/5}$ and $U^{2/3}$ of the hydrodynamic friction of Sec.~\ref{subsec:smooth}. More precisely, using the Hersey number we can rewrite the scaling as $\mu_{f,\rm large} \sim (\eta U/F_N)^{2/5}\bar{F}_N^{1/10}$ at large deformations, and $\mu_{f,\rm small} \sim (\eta U/F_N)^{2/3}\bar{F}_N^{-1/6}$ at small deformations. Hence, an explicit $\bar{F}_N$ dependency remains on top of $\eta U/F_N$, which explains the lack of perfect collapse of the data in the hydrodynamic lubrication regime in Fig.~\ref{fig7}(a).

To overcome this limitation, Fig.~\ref{fig7}(b) presents an alternative Stribeck-like representation, where the friction coefficient is plotted against the theoretical hydrodynamic friction $\mu_{f,\mathrm{th}}(\lambda)$ from \eqref{eq:interpolated_hydrodynamic_friction}. In this form, all curves collapse onto the $y=x$ line in the hydrodynamic regime, independent of the imposed load. Since $\mu_{f,\mathrm{th}}(\lambda)$ is uniquely determined by the dimensionless velocity $\lambda$, this representation contains the same information as a plot $\mu$ versus $\lambda$.

Figure~\ref{fig7}(c) shows the experimental Stribeck curves reported by Bongaerts et al.~\cite{bongaerts2007soft} for PDMS/PDMS contacts with different surface roughnesses. The experimental results highlight that Stribeck curves strongly depend on $h_\mathrm{rms}$, in qualitative agreement with our numerical predictions. It should be noted, however, that the experiments were performed with a sphere sliding against a flat disk, whereas our model assumes a contact between a infinitely long cylinder and a flat surface. A direct quantitative comparison is therefore not possible, although our numerical framework could, in principle, be extended to spherical geometries.

\subsection{Numerical profiles along the transitions}

\begin{figure}[h!]
\centerline{\includegraphics{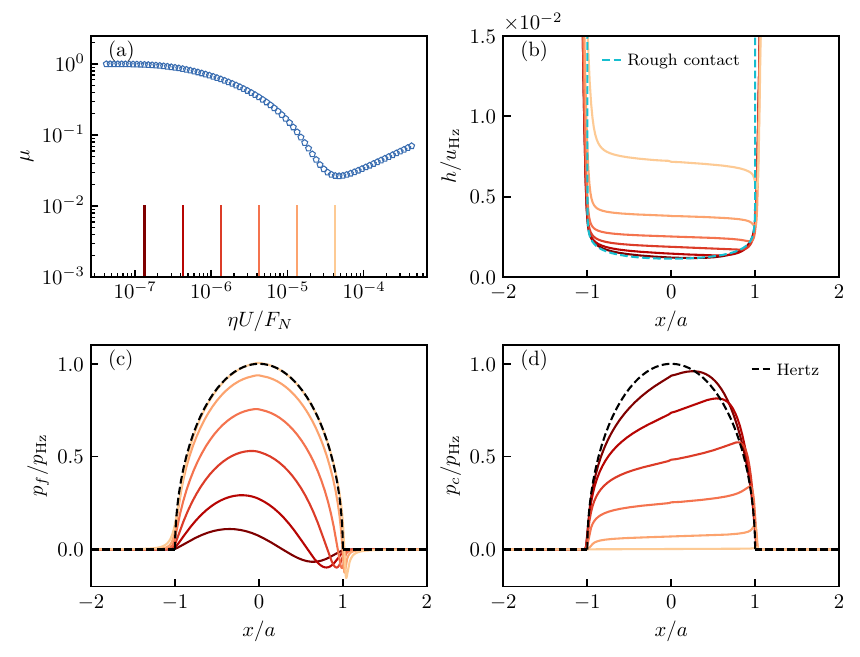}}	
\caption{\label{fig8} \textbf{Profiles in the mixed-lubrication regime.} (a) Stribeck curve with vertical colored lines marking the values of $\eta U/F_N$ corresponding to the profiles shown below. (b) Separation distance, (c) fluid pressure, and (d) contact pressure as a function of the lateral position for various $\lambda$. The cyan and black dashed lines indicate, respectively, the rough-contact asymptotic expression for the separation distance and the Hertzian pressure profile.}
\end{figure}

We now turn to the transition region between the different friction regimes. To gain physical insights, we examine representative profiles of the separation distance, fluid pressure, and contact pressure, shown in Fig.~\ref{fig8}(b)–(d) for several values of the dimensionless sliding speed $\lambda$ in the range $[3.16 \times 10^{-6}, 10^{-3}]$. The contact parameters are fixed to $\bar{F}_N = 0.1$ and $\bar{h}_\mathrm{rms} = 10^{-4}$, corresponding to the blue circles in Fig.\ref{fig7}. The Stribeck curve in Fig.~\ref{fig8}(a) indicates the position of each profile with vertical colored lines. As $\lambda$ increases, the separation distance and fluid pressure gradually rise, while the contact pressure decreases. The rough-contact asymptotic expression of Eq.~\eqref{eq:central_region_contact} (cyan dashed lines) describes the separation distance well at low speed. We observe that the contact pressure converges towards the Hertzian prediction Eq.~\eqref{eq:Hertz} in the boundary lubrication limit (dashed lines in Fig.~\ref{fig8}(d)) since the chosen parameters $\bar{h}_\mathrm{rms} / \bar{F}_N \ll 1$ lies in the large deformation limit (see Sec.~\ref{subsubsec:large_quasi}). Similarly, the fluid pressure profile at the transition to hydrodynamic (Fig.~\ref{fig8}) converges to the same underlying Hertzian prediction, as it balances elastic deformations. 

Overall, the asymptotic solutions of Secs.~\ref{subsec:smooth} and \ref{subsec:quasistatic} accurately capture the separation distance profile in the hydrodynamic and boundary limits. The mixed-lubrication regime, however, does not exhibit a sharp boundary, but rather a smooth crossover between the two extremes as $\lambda$ increases. To obtain predictive descriptions of this intermediate regime, we therefore develop regular perturbative expansions around the hydrodynamic-dominated state (Sec.~\ref{subsec:hydro-to-mixed}) and the contact-dominated state (Sec.~\ref{subsec:contact-to-mixed}).

\subsection{Hydrodynamic-to-mixed lubrication transition}
\label{subsec:hydro-to-mixed}

We begin with the hydrodynamic-to-mixed lubrication transition. Our central assumption here is that the contact pressure remains much smaller than the fluid pressure, so that, to leading order, the separation distance profile is well described by the smooth-limit solution. Under this assumption, the effective contact pressure can be evaluated directly from \eqref{eq:contact_pressure}, and integrated to obtain the contact contribution to the overall friction. Accordingly, the total friction is written as
\begin{equation}
\label{eq:friction_hydro-mixed_transition1}
\mu = \mu_c[h_f(x)] + \mu_{f, \rm th},
\end{equation}
where $\mu_{f, \rm th}$ is given in \eqref{eq:interpolated_hydrodynamic_friction} and 
\begin{equation}
\label{eq:contact_pressure_integral}
\mu_c[h_f(x)] = \frac{\mu_0}{F_N}  \int_\mathbb{R} p_c(x) \, \mathrm{d}x = \frac{\mu_0}{F_N}  \int_\mathbb{R} \beta E^* \exp\left[-\frac{\alpha h_f(x)}{h_\mathrm{rms}}\right] \, \mathrm{d}x,
\end{equation}
with $h_f(x)$ the smooth-limit separation profile. 

We proceed to obtain a simple asymptotic expression for \eqref{eq:friction_hydro-mixed_transition1}. Provided the roughness parameter $\bar{h}_\mathrm{rms}$ is sufficiently small, the transition always occurs for $\lambda \ll 1$, so the hydrodynamic problem falls within the large-deformation ($\lambda \to 0$) regime (see Sec.~\ref{subsubsec:large_deformation_smooth}). In this limit, the integral \eqref{eq:contact_pressure_integral} is dominated by the central region $|x| < a$, where the separation is uniform, $h_{0,\rm large} = 0.44 u_\mathrm{Hz} \lambda^{3/5}$. The fluid contribution may likewise be approximated by its large-indentation form, $\mu_{f,\rm th} \approx \mu_{f,\rm large}$. The total friction then reduces to
\begin{equation}
\label{eq:friction_hydro-mixed_transition}
\mu = \frac{2\mu_0 \beta E^* a}{F_N} \exp\left[-0.44 \frac{\alpha u_\mathrm{Hz}\lambda^{3/5}}{h_\mathrm{rms}} \right] + \frac{\eta U}{F_N} \frac{2a}{0.44 u_\mathrm{Hz} \lambda^{3/5}}.
\end{equation}
In terms of the dimensionless groups introduced earlier, this becomes
\begin{equation}
\label{eq:friction_hydro-mixed_transition_dl}
\mu = \frac{4\mu_0}{\pi} \bar{F}_N^{-1/2} \exp\left[-0.44 \frac{\lambda^{3/5}}{\bar{h}_\mathrm{rms}/\bar{F}_N} \right] + \frac{4\beta}{0.44 \times 3\pi} \lambda^{2/5} \bar{F}_N^{1/2}.
\end{equation}
\begin{figure}[h!]
\centerline{\includegraphics{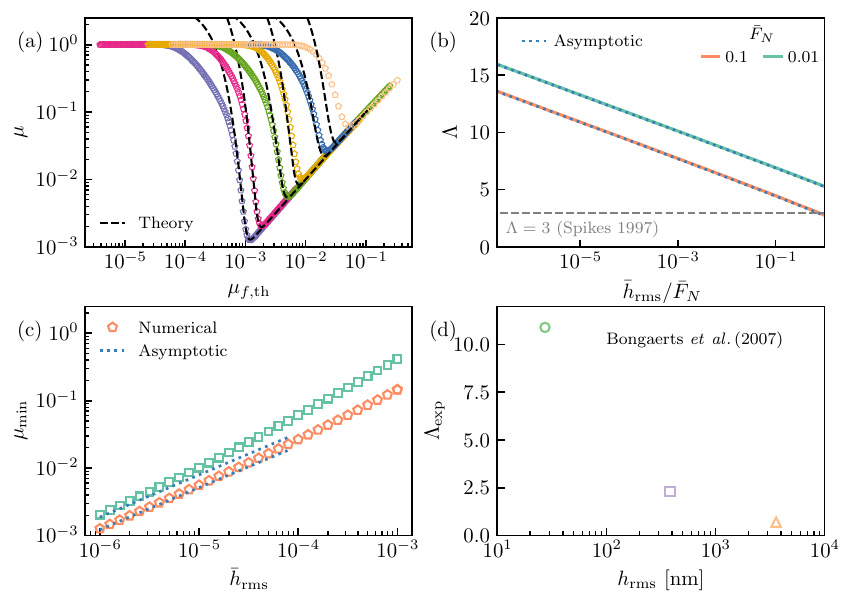}}	
\caption{\label{fig9} \textbf{Hydrodynamic-to-mixed lubrication transition.} (a) Friction coefficient versus the fluid friction as in Fig.~\ref{fig7}(b), where the black dashed lines show the prediction of \eqref{eq:friction_hydro-mixed_transition_dl}, describing very well the hydrodynamic-to-mixed lubrication transition. (b) Dimensionless ratio between the typical film thickness and roughness $\Lambda$ at the minimal friction (see \eqref{eq:Lambda_solution}) versus the dimensionless roughness. The different colors correspond to various normalized indentation force and the dotted lines show the asymptotic expression~\eqref{eq:Lambda_asymptotic}. (c) Numerical (dots) and asymptotic (dotted lines) expression of the minimal friction $\mu_\mathrm{min}$ versus the dimensionless roughness. The asymptotic expressions correspond to~\eqref{eq:minimal_friction}. (d) Experimental value of $\Lambda$ versus the rms roughness in Ref.~\cite{bongaerts2007soft}. }
\end{figure}
Figure~\ref{fig9}(a) shows that the asymptotic expression \eqref{eq:friction_hydro-mixed_transition_dl} (black dashed lines) matches the numerical data closely in the vicinity of the minimal friction and over much of the mixed regime. Deviations appear for large values of $\bar{h}_\mathrm{rms}/\bar{F}_N$ (beige circles), where the transition occurs at $\lambda \sim 1$ and the hydrodynamic contribution is no longer in the large-indentation limit. An interpolation between the two hydrodynamic regimes could improve the prediction (as in Sec.~\ref{subsubsec:general}), but here we focus on the asymptotic case $\bar{h}_\mathrm{rms}/\bar{F}_N \ll 1$ in order to obtain analytic predictions for the minimal friction.

A common way to characterize the hydrodynamic-to-mixed transition is through the $\Lambda$-ratio of film thickness to roughness. It is often assumed that the transition occurs when the film thickness is about three times the roughness scale~\cite{spikes1997mixed}. However, experiments show a much wider range, with reported values from $\Lambda \simeq 0.7–11$~\cite{bongaerts2007soft}, $\Lambda \simeq 0.7–1.2$ \cite{petrova2019fluorescence}, $\Lambda \simeq 0.9–2.3$ \cite{sadowski2019friction}, and $\Lambda \simeq 16–2{,}736$ \cite{dong2023transition}. 
These discrepancies suggest that the transition cannot, in general, be characterized by a universal critical value of $\Lambda$, although roughness-induced contact remains an important transition mechanism in many systems.
In the present asymptotic regime ($\bar{h}_\mathrm{rms}/\bar{F}_N \ll 1$), the transition lies in the large-deformation hydrodynamic limit, leading us to define
\begin{equation}
\Lambda = \frac{h_{0,\mathrm{large}}}{(h_\mathrm{rms}/\alpha)}=\frac{0.44 \lambda_\mathrm{min}^{3/5}}{\bar{h}_\mathrm{rms}/\bar{F}_N},
\end{equation}
where $\lambda_\mathrm{min}$ is the value of $\lambda$ at which friction is minimal, defined explicitly by $\left.\frac{\mathrm{d}\mu}{\mathrm{d}\lambda}\right|_{\lambda=\lambda_\mathrm{min}}=0$. Differentiating \eqref{eq:friction_hydro-mixed_transition_dl} yields the implicit relation
\begin{equation}
\label{eq:Lambda_solution}
\Lambda^{1/3} \exp\left(-\Lambda\right) = k, \quad \quad \text{with} \, \, k = \frac{2}{9 \times (0.44)^{5/3}}\frac{\beta \bar{h}_\mathrm{rms}^{2/3}\bar{F}_N^{1/3}}{\mu_0}. 
\end{equation}
This equation can be inverted using the $-1$ branch of the Lambert W-function~\cite{corless1996lambert}, giving $\Lambda = -\frac{1}{3}W_{-1}\left(-3k^{3}\right)$. In Fig.~\ref{fig9}(b), we plot with solid lines the resulting analytic solution~\eqref{eq:Lambda_solution} of $\Lambda$ versus the dimensionless roughness for two imposed normal forces. We notice that we displayed two curves to illustrate the dependency on both parameters, even though $\Lambda$ is actually a function of a single reduced parameter: $\bar{h}_\mathrm{rms}^{2/3}\bar{F}_N^{1/3}$. A logarithmic scale is used in $x$-axis, highlighting that $\Lambda$ has a typical logarithmic dependency in the contact parameters $\bar{h}_\mathrm{rms}$ and $\bar{F}_N$, which contrasts with other theories assuming a constant $\Lambda = 3$ at the transition as in Ref.~\cite{spikes1997mixed} (e.g. gray dashed lines in Fig.~\ref{fig9}(b)). In the small-roughness limit ($k \ll 1$), expansion of \eqref{eq:Lambda_solution} yields
\begin{equation}
\label{eq:Lambda_asymptotic}
\Lambda \simeq - \ln k + \tfrac{1}{3}\ln[-\ln k],
\end{equation}
which matches the full solution remarkably well across the explored range [Fig.~\ref{fig9}(b), dotted lines].

These results are qualitatively consistent with the experiments of Bongaerts et al.~\cite{bongaerts2007soft}, who reported that $\Lambda$ increases as the surface roughness decreases. In Fig.~\ref{fig9}(d), we plot the experimental values of $\Lambda$ against $h_\mathrm{rms}$ using the same dataset as in Fig.~\ref{fig7}(c). As before, a direct quantitative comparison cannot be made because their measurements were performed with a sphere-on-flat configuration, whereas our model assumes an infinitely long cylinder.

With $\Lambda$ in hand, we can now obtain the minimal friction. The corresponding $\lambda$-value is
\begin{equation}
\label{eq:lambda_min}
\lambda_\mathrm{min} = \left(\frac{\bar{h}_\mathrm{rms}(-\ln k)}{0.44 \bar{F}_N}\right)^{5/3},
\end{equation}
where the sub-logarithmic corrections $\propto \ln[-\ln k]$ in \eqref{eq:Lambda_asymptotic} have been neglected for clarity. 
Inserting $\lambda_\mathrm{min}$ into \eqref{eq:friction_hydro-mixed_transition_dl} then yields the asymptotic minimal friction,
\begin{equation}
\label{eq:minimal_friction}
\mu_\mathrm{min} \simeq \frac{4\beta\bar{h}_\mathrm{rms}^{2/3}\bar{F}^{-1/6}_N}{3\pi \times (0.44)^{5/3}} \bigg[\frac{2}{3}(-\ln k)^{-1/3} + (-\ln k)^{-2/3} + \mathcal{O}[\ln (-\ln k)]\bigg].
\end{equation}
The numerically-extracted value of the minimal friction is shown in Fig.~\ref{fig9}(c), and again compares well with the asymptotic expression for $\bar{h}_\mathrm{rms}$. We recover the trends observed numerically in Fig.~\ref{fig7}, that the overall minimal friction decreases with decreasing roughness $\bar{h}_\mathrm{rms}$ and increasing normal force $\bar{F}_N$.

To summarize, we performed an asymptotic expansion of the total friction around the hydrodynamic smooth limit, which turns out to provide a good description of the mixed-lubrication regime. We focus on the $\bar{h}_\mathrm{rms} \to 0$ limit and derived an asymptotic expression for the $\Lambda$-ratio \eqref{eq:Lambda_asymptotic}, as well as the overall minimal friction \eqref{eq:minimal_friction}. Importantly, we found that $\Lambda$ is not constant but increases logarithmically with decreasing $h_\mathrm{rms}$. 
The resulting criterion predicts transition parameters of the correct order of magnitude and reproduces the qualitative dependence on the roughness amplitude reported by Bongaerts \emph{et al.}~\cite{bongaerts2007soft}.
However, our expression of the mixed-lubrication friction fails at describing the $\lambda \to 0$ limit, hence the transition to the boundary lubrication regime. In the following section, we aim at deriving a criterion defining this transition.

\subsection{Boundary-to-mixed lubrication transition}
\label{subsec:contact-to-mixed}

We now turn to the near-contact regime, corresponding to boundary lubrication ($\lambda \to 0$). As before, we focus on the $\bar{h}_\mathrm{rms} \to 0$ limit, which corresponds to the large-indentation case of Sec.~\ref{subsec:quasistatic}. In this regime, the separation distance is described to leading order by the rough-contact solution \eqref{eq:central_region_contact}. Integrating the resulting hydrodynamic friction via \eqref{eq:friction} yields a fluid friction that scales as $\mu_f \sim \eta U a /(F_N h_\mathrm{rms}) \propto \lambda$, with logarithmic corrections. This contribution is negligible compared to the Coulomb friction $\mu_0$ as $\lambda \to 0$. The balance between the fluid and Coulomb friction leads to critical speeds that are way too large compare to the relevant ones characterizing the boundary-to-mixed transition (not shown explicitly). Therefore the criterion must invoke other arguments.

Another possible scenario for the the boundary-to-mixed lubrication transition is that it is governed by the redistribution of the load between contact and fluid pressures. Let $F_f$ denotes the part of the total load supported by the fluid,
\begin{equation}
F_f = \int_\mathbb{R} p_f(x), \mathrm{d}x.
\end{equation}
Neglecting the fluid friction term in \eqref{eq:friction} and using \eqref{eq:mean-field}, the total friction simplifies to
\begin{equation}
\label{eq:friction_contact}
\mu = \mu_0 \int_\mathbb{R} \frac{p_c(x)}{F_N} \mathrm{d}x = \mu_0 \left(1 - \frac{F_f}{F_N}\right).
\end{equation}
Thus, the reduction of Coulomb friction is directly proportional to the fraction of load carried by the fluid. Hence, our objective in this section is to determine the fluid normal force $F_f$ in the $\lambda \to 0$ limit.

To this end, we expand all fields in $\lambda$ to second order:
\begin{subequations}
\begin{equation}
h = h_0(x) + \lambda h_1(x), \quad \quad h^* =  h_0^* + \lambda h_1^*,
\end{equation}
\begin{equation}
p_f(x) = \lambda p_{f,1}(x) + \lambda^2 p_{f,2}(x), \quad \quad p_c = p_0(x) - \bigg[\lambda p_{f,1}(x) + \lambda^2 p_{f,2}(x)\bigg],
\end{equation}
\end{subequations}
where the leading-order solutions $h_0$ and $p_0$ are the quasi-static solutions discussed in Sec.~\ref{subsec:quasistatic}. The logarithmic divergence of $h_0(|x| \to a^-) \sim -\ln(a-|x|)$ near the contact edge is integrable, and thus does not affect higher-order terms. The detailed resolution of each term of the expansion is provided in Appendix~\ref{app:perturbation}. The first-order fluid pressure $p_{f,1}$ is antisymmetric, such that it does not contribute to the fluid normal force. However, the second-order fluid pressure $p_{f,2}$ has a net normal contribution, which can be integrated to find the fluid normal force, which takes the form
\begin{equation}
\label{eq:fluid-normal}
\frac{F_f}{F_N} = \left(\frac{6\pi \eta U a}{p_\mathrm{Hz}(h_\mathrm{rms}/\alpha)^2}\right)^2 \phi\left(\bar{F}_N\right),
\end{equation}
where we introduce $\phi$ as an auxiliary function of $\bar{F}_N$. Interestingly, the key dimensionless parameter of transition is a (squared) ratio between a lubrication pressure built on the typical roughness scale $6\eta Ua/h_\mathrm{rms}^2$ over the confining Hertz pressure.

\begin{figure}[h!]
\centerline{\includegraphics{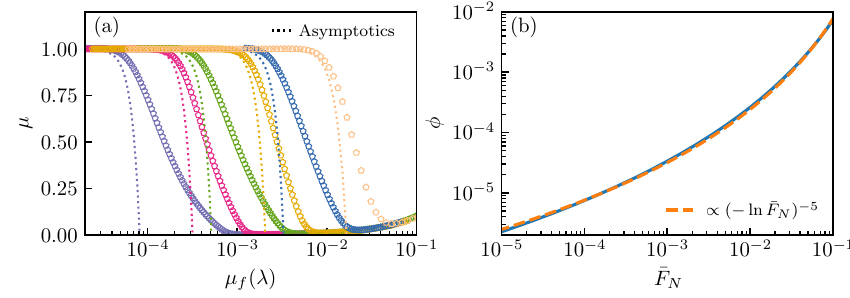}}	
\caption{\label{fig10} \textbf{Boundary-to-mixed lubrication transition.} (a) Friction coefficient versus the fluid friction as in Fig.~\ref{fig7}(b), where the colored dotted lines show the asymptotic expression \eqref{eq:friction_contact} together with \eqref{eq:fluid-normal}. We notice that the $y$-axis is linear here in contrast with previous Stribeck curves of the manuscript. (b) Auxiliary function $\phi$ versus $\bar{F}_N$. The dashed line shows the expression~\eqref{eq:phi_approximation}, which provides a very good approximation of $\phi$.}
\end{figure}

Figure~\ref{fig10}(a) shows the resulting prediction for the friction coefficient, combining \eqref{eq:friction_contact} with \eqref{eq:fluid-normal} (dotted lines). The asymptotic description matches the numerical data remarkably well in the $\lambda \to 0$ limit. Based on this, we propose a criterion for the boundary-to-mixed transition: the value of $\lambda$ at which the fluid load $F_f$ becomes comparable to the imposed normal load, i.e.
\begin{equation}
\label{eq:boundary-to-mixed}
\frac{\lambda^2\phi}{(\bar{h}_\mathrm{rms}/\bar{F}_N)^4}= \left(\frac{6\eta Ua}{p_\mathrm{Hz}(h_\mathrm{rms}/\alpha)^2}\right)^2 \phi\bigg(\bar{F}_N\bigg) = 1.
\end{equation}
Graphically, this corresponds to the intercept of the asymptotic curve in Fig.~\ref{fig10}(a) with the $y=0$ axis, providing a reliable approximation of the transition point.

Lastly, our prediction involves an auxiliary function $\phi$, which we now aim at providing an approximate expression. We plot $\phi$ versus $\bar{F}_N$ in Fig.~\ref{fig10}(b) with logarithmic scales. We observe that $\phi$ increases with $\bar{F}_N$ without any apparent clear scaling. Investigating closely the perturbation analysis, the function $\phi$ depends on the logarithm of $\bar{F}_N$. As logarithm is a slowly varying function, one would naively expect a relatively slow variation of $\phi$ with $\bar{F}_N$. Nevertheless, if one performs a $\ln \bar{F}_N \ll 1$ expansion of $\phi$, it turns out that the leading-order term is $(-\ln \bar{F}_N)^5$ (see Appendix~\ref{app:small_indentation}), ultimately yielding significant variations. In Fig.~\ref{fig10}(b), we show in dashed lines the approximation
\begin{equation}
\label{eq:phi_approximation}
\phi(\bar{F}_N) \approx \frac{1}{2} \, \left(-\ln \bar{F}_N\right)^5
\end{equation}
which provides a very good description of $\phi$ over the entire range explored here and allows us to get a simple closed-form solution for the boundary-to-mixed transition. 
A least-squares fit to the numerical solution yields the asymptotic approximation $\phi(\bar{F}_N) \approx 0.510095 \, \left(-\ln \bar{F}_N\right)^5$. Since the prefactor is remarkably close to 1/2, we use the simpler approximation given by Eq.~\eqref{eq:phi_approximation} throughout the remainder of the paper.

In summary, the boundary-to-mixed transition is controlled by the fraction of the load supported by the fluid pressure. A perturbative $\lambda \to 0$ expansion leads to the criterion $$\lambda = \frac{ (\bar{h}_\mathrm{rms}/\bar{F}_N)^2}{\sqrt{1/2(-\ln \bar{F}_N})^5}$$ providing a prediction for the onset of mixed lubrication.

\section{Conclusion and outlook}
\label{sec:conclusion}
\begin{figure}[h!]
\centerline{\includegraphics{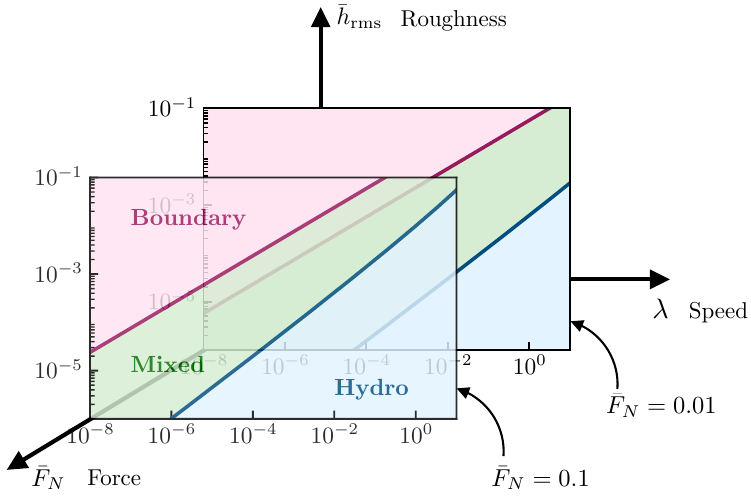}}	
\caption{\label{fig11} \textbf{3d Stribeck curve.} The frictional state is indicated with colored regions in a 3d phase map following the representation of Fig.~\ref{fig3}(b). More precisely, the lines separating the domains correspond to \eqref{eq:speed_boundary-to-mixed} and \eqref{eq:speed_mixed-to-hydro}. The dimensionless normal force are set to $\bar{F}_N = 0.1$ (resp. $0.01$) in the front (resp. back). }
\end{figure}

This work provides a unified framework to characterize the transitions between lubrication regimes along the Stribeck curve. Building on the mean-field theory of Persson and Scaraggi~\citep{persson2009transition}, we first identified three independent dimensionless parameters governing the frictional state, which can be interpreted as a dimensionless speed, load, and roughness.

We then analyzed the hydrodynamic and boundary lubrication regimes (Sec.~\ref{sec:limiting_cases}), which emerge naturally as the smooth-surface limit $h_\mathrm{rms}\to 0$ and the quasistatic limit $U\to 0$ of the present model. We showed that each regime is primarily controlled by a single dimensionless number, namely $\lambda$ and $\bar{h}_\mathrm{rms}/\bar{F}_N$, respectively, and that the contact morphology displays remarkably similar phenomenology as these parameters are varied. In the limits where both parameters are small compared to unity, elastic deformations are large relative to the separation distance, and the contact morphology converges toward the Hertzian solution for dry contact, with non-trivial regularizing boundary layers at the edge of the contact zone. In contrast, for large values of $\lambda$ and $\bar{h}_\mathrm{rms}/\bar{F}_N$, the separation distance are large compare with elastic deformations, and the film thickness can be obtained through a regular perturbation expansion. In the hydrodynamic lubrication regime, we further proposed a prediction for the fluid friction based on an interpolation between the exact asymptotic expressions in the small- and large-deformation limits.

We then turned to the open problem of mixed lubrication and derived asymptotic expressions for the friction coefficient approaching hydrodynamic lubrication and approaching boundary lubrication. In the first case (mixed-to-hydro), the total friction can be accurately described as the sum of the conventional hydrodynamic (viscous) friction and an additional contribution due to residual asperity contacts, leading to
\begin{equation}
\mu = \frac{\mu_0 E^* \ell}{F_N} \exp \left( -\frac{h_\mathrm{fluid}}{h_\mathrm{rms}}\right) + \frac{\eta U\ell }{F_N h_\mathrm{fluid}},
\end{equation}
where $\ell$ and $h_\mathrm{fluid}$ denote the characteristic contact length and fluid film thickness, as predicted by EHD lubrication theory. For sufficiently smooth surfaces, i.e. $h_\mathrm{rms}/R \ll 1$, the transition occurs in the large-deformation limit where $\ell = 2a$ and $h_\mathrm{fluid} = 0.44 u_\mathrm{Hz} \lambda^{3/5}$ . In this regime, we derived an explicit expression for the film-thickness-to-roughness ratio (the $\Lambda$-ratio) at the transition,
\begin{equation}
\Lambda \approx -\ln k + \frac{1}{3}\ln \left[-\ln k\right], \quad \quad k = \frac{2}{9 \times (0.44)^{5/3}}\frac{\beta }{\mu_0} \left(\frac{h_\mathrm{rms}^2F_N}{16\pi \alpha^2 \beta^6 E^*R^3}\right)^{1/3},
\end{equation}
which contrasts with approaches assuming a constant critical value of $\Lambda$~\cite{spikes1997mixed}. In particular, this result recovers the trend reported by Bongaerts \textit{et al.}, namely that $\Lambda$ increases as the surface roughness decreases~\cite{bongaerts2007soft}. This criterion can be translated into a critical sliding speed,
\begin{equation}
\label{eq:speed_mixed-to-hydro}
U = \frac{p_\mathrm{Hz}u_\mathrm{Hz}^2}{6\eta a }\left(\frac{-h_\mathrm{rms}\ln k}{0.44\alpha u_\mathrm{Hz}}\right)^{5/3} = \frac{4F_N^2}{3\pi^2 E^* R \eta  }\left(\frac{-\pi E^* h_\mathrm{rms}\ln k}{4 \times 0.44\alpha F_N}\right)^{5/3}.
\end{equation}
Determining this critical speed further allows one to obtain an expression for the global minimum of the friction coefficient, which scales as $h_\mathrm{rms}^{2/3}F_N^{-1/6}$, with logarithmic corrections.

In the second transition of boundary-to-mixed lubrication, the viscous dissipation in the fluid is negligible, and the reduction of friction is instead controlled by the progressive redistribution of the applied load between direct solid–solid contact and hydrodynamic pressure. In this regime, the total friction coefficient admits the asymptotic form
\begin{equation}
\mu = \mu_0 \left(1 - \frac{F_f}{F_N}\right), \quad \quad \frac{F_f}{F_N}= 0.5 \left(\frac{12\pi \eta R U }{E^* (h_\mathrm{rms}/\alpha)} \right)^2 \, \left(\ln \left[ \frac{1}{\pi \beta^2}\frac{F_N}{R E^*}\right]\right)^{5}
\end{equation}
where $F_f$ denotes the amount of the normal load supported by the fluid. From this result, one can define a characteristic transition speed at which the load carried by the fluid becomes significant,
\begin{equation}
\label{eq:speed_boundary-to-mixed}
U = \frac{p_\mathrm{Hz}h_\mathrm{rms}^2}{6\eta a \alpha^2} \left(-0.5\ln \left[ \frac{1}{\pi \beta^2}\frac{F_N}{R E^*}\right]\right)^{-5/2} = \frac{E^*h_\mathrm{rms}^2}{12\eta R \alpha^2} \left(-0.5\ln \left[ \frac{1}{\pi \beta^2}\frac{F_N}{R E^*}\right]\right)^{-5/2}.
\end{equation}

The resulting phase diagram (Fig.~\ref{fig11}), expressed in the plane of dimensionless speed and dimensionless roughness, summarizes the different lubrication regimes identified in this work. At fixed roughness, increasing the sliding speed drives the system successively from boundary to mixed and finally to hydrodynamic lubrication. Conversely, at fixed speed, decreasing the surface roughness induces the same sequence of transitions. 
The sequence of three lubrication regimes described above persists only while the boundary-to-mixed and mixed-to-hydrodynamic transition lines remain distinct. For sufficiently large roughness amplitudes, the minimum friction predicted in the mixed-lubrication regime exceeds the boundary friction coefficient $\mu_0$. In that case, the boundary-to-mixed and mixed-to-hydrodynamic transition lines merge, resulting in a direct transition from boundary to hydrodynamic lubrication. In this case, the friction coefficient varies monotonically with sliding velocity, and the velocity-weakening branch disappears.
A key conclusion of this analysis is that the classical notion of the Stribeck curve can be naturally generalized to a multidimensional framework, provided that appropriate dimensionless parameters are used to capture the microscopic properties of the contacting surfaces. The present framework is restricted to statistically rough surfaces described by Persson's contact theory. An interesting extension would be to investigate how deterministic surface textures modify the lubrication transitions and the associated phase diagram~\cite{choo2008,peng2021bending,peng2021elastohydrodynamic,kargar2025non,minten2026hydrodynamic}.

More broadly, our work demonstrates that asymptotic expansions provide a systematic and rigorous route to characterize transitions between frictional states. At the same time, the present mean-field lubrication model relies on a number of simplifying assumptions. In particular, it neglects various effects like flow factors~\cite{patir1978}, possible percolation effects in the contact network that may entirely block the fluid flow~\cite{persson2011lubricated}, non-Newtonian rheology in nanometric fluid films~\cite{jadhao2019rheological}, as well as intermolecular interactions~\cite{israelachvili2011intermolecular,dong2025role}, and possible cavitation at the outlet~\cite{richards2023anomalous}. Accounting for these effects may be essential in specific situations and could modify both the location and nature of the predicted transitions. We restrict ourselves to an infinite cylinder for the sake of simplicity, but the framework developed here could be extended to spherical contacts. More generally, the framework developed here may have implications across a broad range of systems, from classical tribological contacts to the rheology of complex suspensions, where frictional transitions at the grain scale are known to strongly influence macroscopic flow properties~\cite{guazzelli2018rheology,lobry2019shear,ness2022physics}.

\section*{Acknowledgements}
This work has received funding from the European Research Council (ERC) under the European Union’s Horizon (Grant 101097842, CohPa). Views and opinions expressed are, however, those of the author(s) only and do not necessarily reflect those of the European Union or the European Research Council. Neither the European Union nor the granting authority can be held responsible for them.

\appendix
\section{Dimensionless system of equations}
\label{app:dimensionless}
In this section, we write the dimensionless system of equations that has been solved. We introduce dimensionless variables with bar, as $\bar{h} = h/u_\mathrm{Hz}$, $\bar{p} = p/p_\mathrm{Hz}$ and so on, such that the system of equations \eqref{eq:mean-field} - \eqref{eq:elasticity} becomes
\begin{equation}
\label{eq:mean-field_dl}
\bar{p}(\bar{x}) = \bar{p}_f(\bar{x}) + \bar{p}_c(\bar{x}), 
\end{equation}
\begin{equation}
\label{eq:load_dl}
\int_\mathbb{R} \bar{p}(\bar{x}) \, \mathrm{d}\bar{x} = \frac{\pi}{2},
\end{equation}
\begin{equation}
\label{eq:contact_pressure_dl}
\bar{p}_c(\bar{x}) = \bar{F}_N^{-1/2} \exp\left(- \frac{\bar{h}(\bar{x})}{\bar{h}_\textrm{rms} /\bar{F}_N}\right),
\end{equation}
\begin{equation}
\label{eq:thin-film_dl}
\bar{p}_f'(\bar{x}) = \lambda \frac{\bar{h}(\bar{x}) - \bar{h}^*}{\bar{h}^3(\bar{x})},
\end{equation}
\begin{equation}
\label{eq:elasticity_dl}
\bar{h}(\bar{x}) = \bar{z}_0 + \frac{\bar{x}^2}{2} - \frac{1}{\pi} \int_\mathbb{R} \bar{p}(\bar{y}) \ln \vert \bar{x}-\bar{y}\vert\, \mathrm{d}\bar{y}.
\end{equation}
We solve the latter system of equations numerically. The derivative in the thin-film equation~\eqref{eq:thin-film_dl} is discretized with a finite-difference scheme. We use a standard continuation algorithm to obtain  solutions varying independently the three parameters.

\section{Boundary-layer structure in the quasistatic limit}
\label{app:boundarylayer}

In this appendix, we rationalize the boundary layer observed at the edge of the contact zone in the quasistatic limit ($\lambda = 0$), as shown in Fig.~\ref{fig6}(c--e). In this limit, the fluid pressure vanishes ($p_f=0$) and the problem reduces to the coupled contact and elastic equations \eqref{eq:contact_pressure_dl} and \eqref{eq:elasticity_dl}, with $p = p_c$.

\vspace{0.5em}
\noindent
\textit{Physical picture.} — The rough contact solution derived in the main text provides an accurate description of the separation profile in the bulk of the contact region. However, it exhibits a logarithmic divergence as $\lvert x \rvert \to a^{-}$, while the Hertz solution outside the contact behaves as $h \sim (x-a)^{3/2}$. These two asymptotics are incompatible, which necessarily implies the existence of a narrow transition region near $x \approx \pm a$. This region plays the role of a boundary layer that regularizes the logarithmic singularity and connects smoothly the inner (rough contact) and outer (Hertzian) solutions.

\vspace{0.5em}
\noindent
\textit{Scaling analysis.} — We denote by $\hat{\ell}$, $\hat{h}$ and $\hat{p}$ the characteristic lateral extent, separation distance, and pressure within this boundary layer (see Fig.~\ref{fig6}(e)). Inside the contact region, the separation is controlled by roughness and scales as
\begin{equation}
\hat{h} \sim \frac{h_\mathrm{rms}}{\alpha}.
\end{equation}
To estimate the pressure scale, we differentiate twice the elasticity equation \eqref{eq:elasticity_dl}, yielding
\begin{equation}
\bar{h}''(\bar{x}) = 1 - \frac{1}{\pi} \int_\mathbb{R} \frac{\bar{p}'(\bar{y})}{\lvert \bar{x}-\bar{y} \rvert} \, \mathrm{d}\bar{y}.
\end{equation}
Within the boundary layer, the curvature of the gap is dominated by elastic deformation rather than the sphere geometry, so that $\bar{h}'' \sim \hat{h}/\hat{\ell}^2$. Balancing this with the integral term gives the pressure scale
\begin{equation}
\hat{p} \sim E^* \frac{\hat{h}}{\hat{\ell}}.
\end{equation}
The final ingredient is the matching with the outer Hertz solution. Close to the edge of contact, the Hertz profile behaves as
\begin{equation}
h(x \to a^+) \sim u_\mathrm{Hz} \left(\frac{x-a}{a}\right)^{3/2}.
\end{equation}
Matching this outer behavior with the boundary layer implies
\begin{equation}
\hat{h} \sim u_\mathrm{Hz} \left(\frac{\hat{\ell}}{a}\right)^{3/2}.
\end{equation}
Combining these relations yields the boundary-layer scales
\begin{equation}
\hat{h} = \frac{h_\mathrm{rms}}{\alpha}, 
\qquad
\hat{\ell} = a \left(\frac{h_\mathrm{rms}}{\alpha u_\mathrm{Hz}}\right)^{2/3}, 
\qquad
\hat{p} = p_\mathrm{Hz} \left(\frac{h_\mathrm{rms}}{\alpha u_\mathrm{Hz}}\right)^{1/3}.
\end{equation}

\vspace{0.5em}
\noindent
\textit{Self-similar structure.} — 
These scalings suggest the existence of a universal similarity solution describing the boundary layer. Introducing the rescaled coordinate
\begin{equation}
\xi = \frac{x/a - 1}{(\bar{h}_\mathrm{rms}/\bar{F}_N)^{2/3}},
\end{equation}
and the Ansatz
\begin{equation}
h(x) = \frac{h_\mathrm{rms}}{\alpha} \left[ -\frac{1}{2}\ln \bar{F}_N - \frac{1}{3}\ln\left(\frac{\bar{h}_\mathrm{rms}}{\bar{F}_N}\right) + \mathcal{H}(\xi) \right],
\end{equation}
\begin{equation}
p(x) = p_\mathrm{Hz} \left(\frac{\bar{h}_\mathrm{rms}}{\bar{F}_N}\right)^{1/3} \mathcal{P}(\xi),
\end{equation}
one obtains the closed system
\begin{equation}
\mathcal{P}(\xi) = \exp\left(-\mathcal{H}(\xi)\right), 
\qquad
\mathcal{H}''(\xi) = -\frac{1}{\pi} \, \int_\mathbb{R} \frac{\mathcal{P}'(\bar{\xi})}{\xi - \bar{\xi}} \, \mathrm{d}\bar{\xi},
\end{equation}

\vspace{0.5em}
\noindent
\textit{Numerical validation.} — Figure~\ref{fig6}(c) shows the numerical profiles rescaled according to this similarity form. The excellent collapse confirms that the edge of the contact is governed by a universal boundary-layer solution, which smoothly connects the logarithmic rough-contact solution to the Hertzian outer profile.

\section{Near-contact perturbation analysis}
\label{app:perturbation}
\begin{figure}[h!]
\centerline{\includegraphics{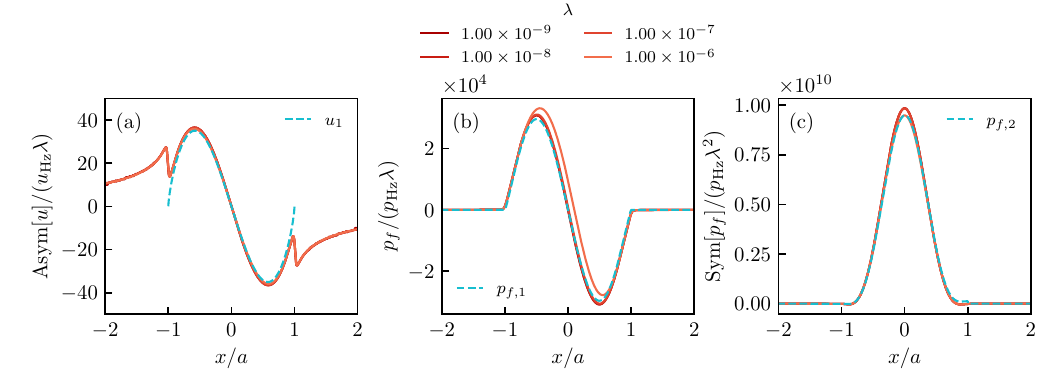}}	
\caption{\label{figS1} \textbf{Small force expansion.}  }
\end{figure}
In this Appendix, we perform the regular perturbation expansion of \eqref{eq:mean-field_dl}-\eqref{eq:elasticity_dl} in $\lambda$, as described in \ref{subsec:contact-to-mixed}. Our goal is to find the part of the force carried by the fluid pressure in the $\lambda \to 0$ limit, which allows to characterize the boundary-to-mixed lubrication transition. We recall the perturbation expansion that we set (in dimensionless form) to
\begin{subequations}
\begin{equation}
\bar{h}(\bar{x}) = \bar{h}_0(\bar{x}) + \lambda \bar{h}_1(\bar{x}), \quad \quad \bar{h}^* =  \bar{h}_0^* + \lambda \bar{h}_1^*,
\end{equation}
\begin{equation}
\bar{p}_f(x) = \lambda \bar{p}_{f,1}(\bar{x}) + \lambda^2 \bar{p}_{f,2}(\bar{x}), \quad \quad \bar{p}_c(\bar{x}) = \bar{p}_0(\bar{x}) - \bigg[\lambda \bar{p}_{f,1}(\bar{x}) + \lambda^2 \bar{p}_{f,2}(\bar{x})\bigg],
\end{equation}
\end{subequations}
where $\bar{h}_0$ and $\bar{p}_0$ are the leading-order quasistatic solution, corresponding to the ones described in Sec.~\ref{subsec:quasistatic}. In dimensionless form, the latter reads
\begin{equation}
\bar{h}_0(\bar{x}) = -\frac{\bar{h}_\mathrm{rms}}{2\bar{F}_N} \ln\left(\bar{F}_N \left(1-\bar x^2\right)\right) \, \quad \quad \bar{p}_0(x) = \sqrt{1-\bar x^2} \Theta(\bar x^2-1).
\end{equation}
At first order $\mathcal{O}(\lambda)$, the thin-film equation \eqref{eq:thin-film_dl} gives 
\begin{equation}
\label{eq:first-order_thin-film}
\bar{p}_{f,1}' = \frac{\bar{h}_0(\bar x) - \bar{h}_0^*}{\bar{h}_0(\bar x)}.
\end{equation}
We continue by integrating the latter equation over the all contact zone and we use the condition $\bar{p}_{f, 1}(\pm \infty) = 0$, in order to find the constant $\bar{h}_0^*$ that reads 
\begin{equation}
\bar{h}_0^* = \frac{\int_{-1}^1 \bar{h}_0^{-2}(\bar x)\,\mathrm{d}\bar x}{\int_{-1}^1 \bar{h}_0^{-3}(\bar x)\,\mathrm{d}\bar x}. 
\end{equation}
Now that we know $\bar h_0^*$, the first-order fluid pressure is obtained integrating \eqref{eq:first-order_thin-film} as
\begin{equation}
\label{eq:pf1}
\bar{p}_{f,1}(\bar{x}) = \int_{-1}^{\bar{x}} \frac{\bar{h}_0(\bar y) - \bar{h}_0^*}{\bar{h}_0(\bar y)}\,\mathrm{d}\bar y.
\end{equation}
In Fig.~\ref{figS1}(a), we plot the fluid pressure rescaled by $\lambda$, which exhibits excellent agreement with the prediction \eqref{eq:pf1}. We notice that we could not find an closed-form expression for $\bar{p}_{f,1}$ with special functions, despite its relative simplicity. Furthermore, we notice that $\bar{p}_{f,1}$ scales with $(\bar{h}_\mathrm{rms}/\bar{F}_N)^{-2}$ but with non-trivial logarithmic dependencies in $\bar{F}_N$. As the leading order separation distance is symmetric, the resulting first-order fluid pressure is antisymmetric, such that it does not contribute to the fluid load. Hence, we must solve for the second order in $\lambda$ to find the fluid normal force.

To do so, we evaluate \eqref{eq:contact_pressure_dl} at the order $\mathcal{O}(\lambda)$ to find the first-order separation distance as
\begin{equation}
\label{eq:h1}
\bar{h}_1(\bar{x}) = \frac{\bar{h}_{\mathrm{rms}}}{\bar{F}_N} \frac{\bar{p}_{f,1}(\bar{x})}{\bar{p}_0(\bar{x})}.
\end{equation}
The Fig.~\ref{fig9}(b) shows the antisymmetric part of the separation distance $\mathrm{Asym}[\bar h](\bar x) = (\bar h(\bar x) - \bar h (-\bar x))/2$ rescaled by $\lambda$. We observe an excellent collapse of the numerical data at small $\lambda$ and $\mathrm{Asym}[\bar h]/\lambda$ is very well described by $\bar h_1$ from \eqref{eq:h1}. We choose to show the antisymmetric part of $\bar h$ instead of $\bar h- \bar h_0$, in order to remove the (symmetric) logarithmic singularity. Nevertheless, the same collapse is observed for $\bar h- \bar h_0$ in the central zone (not shown). Lastly, we notice that the first-order separation $\bar{h}_1$ scales as $\bar{h}_\mathrm{rms}/\bar{F}_N$, again with logarithmic corrections in $\bar{F}_N$.  

Now that we know $\bar h_1$, we can solve for the second-order fluid pressure, by evaluating \eqref{eq:thin-film_dl} at the order $\mathcal{O}(\lambda^2)$, which yields to 
\begin{equation}
\label{eq:thin-film_second-order}
\bar p_{f,2}' = \frac{(3\bar{h}_0^* - 2\bar{h}_0)\bar{h}_1}{\bar{h}_0^4} - \frac{\bar{h}_1^*}{\bar{h}_0^3}. 
\end{equation}
We again integrate \eqref{eq:thin-film_second-order} over the all contact zone and use the boundary condition $\bar{p}_{f,2}(\pm \infty) =0$ to find the constant $\bar{h}_1^*$. The first term of the r.h.s. $(3\bar{h}_0^* - 2\bar{h}_0)\bar{h}_1/\bar{h}_0^4$ of \eqref{eq:thin-film_second-order} is antisymmetric, such that its integration over the all domain vanishes and we simply find the trivial condition $\bar{h}_1^* = 0$. Then, the second-order pressure can be evaluated by integrating \eqref{eq:thin-film_second-order} to get 
\begin{equation}
\label{eq:p2}
\bar{p}_{f,2}(\bar{x}) = \int_{-1}^{\bar x}\frac{(3\bar{h}_0^* - 2\bar{h}_0(\bar{y}))\bar{h}_1(\bar{y})}{ \bar{h}_0^4(\bar{y})} \, \mathrm{d}\bar{y}.
\end{equation}
We show in Fig.~\ref{fig9}(c) the symmetric part of the fluid pressure, which shows excellent agreement with theoretical prediction $\bar{p}_{f,2}$ from \eqref{eq:p2}. Lastly, we note that $\bar{p}_{f,2}$ scales as $(\bar{h}_\mathrm{rms}/\bar{F}_N)^{-4}$ and is symmetric and leads to a positive normal force. The same holds for the fluid normal force, which can be expressed in a dimensionless form as 
\begin{equation}
\label{eq:fluid-normal-dl}
\frac{F_f}{F_N} = \frac{\lambda^2}{(\bar{h}_\mathrm{rms}/\bar{F}_N)^4} \phi(\bar{F}_N)
\end{equation}
where $\phi$ is a dimensionless auxiliary function representing the dependency in the dimensionless normal force.  The equation~\eqref{eq:fluid-normal-dl} corresponds to \eqref{eq:fluid-normal} of the manuscript.  

\section{Small indentation force expansion of the auxiliary function $\phi$}
\label{app:small_indentation}
In this appendix, we further highlight the scaling of the different terms of the small-$\lambda$ expansion in both dimensionless roughness via $\bar{h}_\mathrm{rms}/\bar{F}_N$ and normal force in $\ln \bar{F}_N$. Our objective is to justify the $\phi(\bar{F}_N) \sim (-\ln \bar{F}_N)^{-5}$ scaling discussed in Sec.~\ref{subsec:contact-to-mixed} and Fig.~\ref{fig10}. We go through all the different steps of the perturbation expansion and rewrite the fields in dimensionless form with capital letters, where the dependency in $\bar{h}_\mathrm{rms}/\bar{F}_N$ is scaled out, as  
\begin{subequations}
\begin{equation}
\bar{h}_0(\bar{x})  = \left(\frac{\bar{h}_\mathrm{rms}}{\bar{F}_N}\right) H_0\left(\ln \bar{F}_N, \bar{x}\right), \quad \quad H_0\left(\ln \bar{F}_N, \bar{x}\right) = -\frac{1}{2}\ln \bar{F}_N - \frac{1}{2}\ln(1-\bar{x}^2) \propto -\ln \bar{F}_N,
\end{equation}
\begin{equation}
\bar{h}^*_0  = \left(\frac{\bar{h}_\mathrm{rms}}{\bar{F}_N}\right) H_0^*\left(\ln \bar{F}_N\right), \quad \quad H_0^*\left(\ln \bar{F}_N\right) = \frac{\int_{-1}^1 H_0^{-2}(\ln \bar{F}_N, \bar{x})\,\mathrm{d}\bar{x}}{\int_{-1}^1 H_0^{-3}(\ln \bar{F}_N, \bar{x})\,\mathrm{d}\bar{x}} \propto -\ln \bar{F}_N, 
\end{equation}
\begin{equation}
\bar{p}_{f,1}(\bar{x}) = \left(\frac{\bar{h}_\mathrm{rms}}{\bar{F}_N}\right)^{-2} P_{f,1}(\ln \bar{F}_N, \bar{x}), \quad \quad P_{f,1}(\ln \bar{F}_N, \bar{x}) =\int_{-1}^{\bar{x}} \frac{H_0(\ln \bar{F}_N, \bar y) - H_0^*(\ln \bar{F}_N)}{H_0^3(\ln \bar{F}_N, \bar y)}\,\mathrm{d}\bar y \propto (-\ln \bar{F}_N)^{-2},
\end{equation}
\begin{equation}
\bar{h}_1(\bar{x}) = \left(\frac{\bar{h}_\mathrm{rms}}{\bar{F}_N}\right)^{-1} H_1(\ln \bar{F}_N, \bar{x}), \quad \quad H_1(\ln \bar{F}_N, \bar{x}) = \frac{P_{f,1}(\ln \bar{F}_N, \bar{x})}{p_0(\bar{x})} \propto (-\ln \bar{F}_N)^{-2},
\end{equation}
\begin{equation}
\bar{p}_{f,2}(\bar{x}) = \left(\frac{\bar{h}_\mathrm{rms}}{\bar{F}_N}\right)^{-4} P_{f,2}(\ln \bar{F}_N, \bar{x}), \quad \quad P_{f,2}(\ln \bar{F}_N, \bar{x}) = \int_{-1}^{\bar{x}}\frac{(3H_0^*(\ln \bar{F}_N) - 2H_0(\ln \bar{F}_N, \bar{y}))H_1(\bar{y})}{H_0^4(\ln \bar{F}_N, \bar{y})} \, \mathrm{d}\bar{y}\propto (-\ln \bar{F}_N)^{-5},
\end{equation}
\begin{equation}
\frac{F_f}{F_N}= \lambda^2 \left(\frac{\bar{h}_\mathrm{rms}}{\bar{F}_N}\right)^{-4} \phi \left( \ln \bar{F}_N \right) , \quad \quad \phi\left(\ln \bar{F}_N\right) = \int_{-1}^1P_{f,2}(\ln \bar{F}_N, \bar x) \, \mathrm{d}\bar{x}\propto (-\ln \bar{F}_N)^{-5}.
\end{equation}
\end{subequations}
In the former equations, the $\propto$ correspond to the $\bar{F}_N \ll 1$ asymptotic.

%\bibliography{biblio}

\begin{thebibliography}{56}%
\makeatletter
\providecommand \@ifxundefined [1]{%
 \@ifx{#1\undefined}
}%
\providecommand \@ifnum [1]{%
 \ifnum #1\expandafter \@firstoftwo
 \else \expandafter \@secondoftwo
 \fi
}%
\providecommand \@ifx [1]{%
 \ifx #1\expandafter \@firstoftwo
 \else \expandafter \@secondoftwo
 \fi
}%
\providecommand \natexlab [1]{#1}%
\providecommand \enquote  [1]{``#1''}%
\providecommand \bibnamefont  [1]{#1}%
\providecommand \bibfnamefont [1]{#1}%
\providecommand \citenamefont [1]{#1}%
\providecommand \href@noop [0]{\@secondoftwo}%
\providecommand \href [0]{\begingroup \@sanitize@url \@href}%
\providecommand \@href[1]{\@@startlink{#1}\@@href}%
\providecommand \@@href[1]{\endgroup#1\@@endlink}%
\providecommand \@sanitize@url [0]{\catcode `\\12\catcode `\$12\catcode
  `\&12\catcode `\#12\catcode `\^12\catcode `\_12\catcode `\%12\relax}%
\providecommand \@@startlink[1]{}%
\providecommand \@@endlink[0]{}%
\providecommand \url  [0]{\begingroup\@sanitize@url \@url }%
\providecommand \@url [1]{\endgroup\@href {#1}{\urlprefix }}%
\providecommand \urlprefix  [0]{URL }%
\providecommand \Eprint [0]{\href }%
\providecommand \doibase [0]{http://dx.doi.org/}%
\providecommand \selectlanguage [0]{\@gobble}%
\providecommand \bibinfo  [0]{\@secondoftwo}%
\providecommand \bibfield  [0]{\@secondoftwo}%
\providecommand \translation [1]{[#1]}%
\providecommand \BibitemOpen [0]{}%
\providecommand \bibitemStop [0]{}%
\providecommand \bibitemNoStop [0]{.\EOS\space}%
\providecommand \EOS [0]{\spacefactor3000\relax}%
\providecommand \BibitemShut  [1]{\csname bibitem#1\endcsname}%
\let\auto@bib@innerbib\@empty
%</preamble>
\bibitem [{\citenamefont {Persson}(2013)}]{persson2013sliding}%
  \BibitemOpen
  \bibfield  {author} {\bibinfo {author} {\bibfnamefont {B.~N.}\ \bibnamefont
  {Persson}},\ }\href@noop {} {\emph {\bibinfo {title} {Sliding friction:
  physical principles and applications}}}\ (\bibinfo  {publisher} {Springer
  Science \& Business Media},\ \bibinfo {year} {2013})\BibitemShut {NoStop}%
\bibitem [{\citenamefont {Bhushan}(2001)}]{bhushan2001tribology}%
  \BibitemOpen
  \bibfield  {author} {\bibinfo {author} {\bibfnamefont {B.}~\bibnamefont
  {Bhushan}},\ }\href@noop {} {Tribology on the macroscale to nanoscale of
  microelectromechanical system materials: a review,\ \bibfield  {journal}
  {\bibinfo  {journal} {Proceedings of the Institution of Mechanical Engineers,
  Part J: Journal of Engineering Tribology}\ }\textbf {\bibinfo {volume}
  {215}},\ \bibinfo {pages} {1} (\bibinfo {year} {2001})}\BibitemShut {NoStop}%
\bibitem [{\citenamefont {Moghadasi}\ \emph {et~al.}(2022)\citenamefont
  {Moghadasi}, \citenamefont {Isa}, \citenamefont {Ariffin}, \citenamefont
  {Raja}, \citenamefont {Wu}, \citenamefont {Yamani}, \citenamefont {Muhamad},
  \citenamefont {Yusof}, \citenamefont {Jamaludin}, \citenamefont {bin
  Ab~Karim} \emph {et~al.}}]{moghadasi2022review}%
  \BibitemOpen
  \bibfield  {author} {\bibinfo {author} {\bibfnamefont {K.}~\bibnamefont
  {Moghadasi}}, \bibinfo {author} {\bibfnamefont {M.~S.~M.}\ \bibnamefont
  {Isa}}, \bibinfo {author} {\bibfnamefont {M.~A.}\ \bibnamefont {Ariffin}},
  \bibinfo {author} {\bibfnamefont {S.}~\bibnamefont {Raja}}, \bibinfo {author}
  {\bibfnamefont {B.}~\bibnamefont {Wu}}, \bibinfo {author} {\bibfnamefont
  {M.}~\bibnamefont {Yamani}}, \bibinfo {author} {\bibfnamefont {M.~R.~B.}\
  \bibnamefont {Muhamad}}, \bibinfo {author} {\bibfnamefont {F.}~\bibnamefont
  {Yusof}}, \bibinfo {author} {\bibfnamefont {M.~F.}\ \bibnamefont
  {Jamaludin}}, \bibinfo {author} {\bibfnamefont {M.~S.}\ \bibnamefont {bin
  Ab~Karim}},  \emph {et~al.},\ }\href@noop {} {A review on biomedical implant
  materials and the effect of friction stir based techniques on their
  mechanical and tribological properties,\ \bibfield  {journal} {\bibinfo
  {journal} {Journal of Materials Research and Technology}\ }\textbf {\bibinfo
  {volume} {17}},\ \bibinfo {pages} {1054} (\bibinfo {year}
  {2022})}\BibitemShut {NoStop}%
\bibitem [{\citenamefont {Jahn}\ \emph {et~al.}(2016)\citenamefont {Jahn},
  \citenamefont {Seror},\ and\ \citenamefont {Klein}}]{jahn2016lubrication}%
  \BibitemOpen
  \bibfield  {author} {\bibinfo {author} {\bibfnamefont {S.}~\bibnamefont
  {Jahn}}, \bibinfo {author} {\bibfnamefont {J.}~\bibnamefont {Seror}}, \ and\
  \bibinfo {author} {\bibfnamefont {J.}~\bibnamefont {Klein}},\ }\href@noop {}
  {Lubrication of articular cartilage,\ \bibfield  {journal} {\bibinfo
  {journal} {Annual Review of Biomedical Engineering}\ }\textbf {\bibinfo
  {volume} {18}},\ \bibinfo {pages} {235} (\bibinfo {year} {2016})}\BibitemShut
  {NoStop}%
\bibitem [{\citenamefont {Greenwood}\ and\ \citenamefont
  {Williamson}(1966)}]{greenwood1966contact}%
  \BibitemOpen
  \bibfield  {author} {\bibinfo {author} {\bibfnamefont {J.~A.}\ \bibnamefont
  {Greenwood}}\ and\ \bibinfo {author} {\bibfnamefont {J.~P.}\ \bibnamefont
  {Williamson}},\ }\href@noop {} {Contact of nominally flat surfaces,\
  \bibfield  {journal} {\bibinfo  {journal} {Proceedings of the royal society
  of London. Series A. Mathematical and physical sciences}\ }\textbf {\bibinfo
  {volume} {295}},\ \bibinfo {pages} {300} (\bibinfo {year}
  {1966})}\BibitemShut {NoStop}%
\bibitem [{\citenamefont {Persson}(2001)}]{persson2001theory}%
  \BibitemOpen
  \bibfield  {author} {\bibinfo {author} {\bibfnamefont {B.~N.}\ \bibnamefont
  {Persson}},\ }\href@noop {} {Theory of rubber friction and contact
  mechanics,\ \bibfield  {journal} {\bibinfo  {journal} {The Journal of
  Chemical Physics}\ }\textbf {\bibinfo {volume} {115}},\ \bibinfo {pages}
  {3840} (\bibinfo {year} {2001})}\BibitemShut {NoStop}%
\bibitem [{\citenamefont {Gong}(2006)}]{gong2006friction}%
  \BibitemOpen
  \bibfield  {author} {\bibinfo {author} {\bibfnamefont {J.~P.}\ \bibnamefont
  {Gong}},\ }\href@noop {} {Friction and lubrication of hydrogels—its
  richness and complexity,\ \bibfield  {journal} {\bibinfo  {journal} {Soft
  matter}\ }\textbf {\bibinfo {volume} {2}},\ \bibinfo {pages} {544} (\bibinfo
  {year} {2006})}\BibitemShut {NoStop}%
\bibitem [{\citenamefont {Schallamach}(1963)}]{schallamach1963theory}%
  \BibitemOpen
  \bibfield  {author} {\bibinfo {author} {\bibfnamefont {A.}~\bibnamefont
  {Schallamach}},\ }\href@noop {} {A theory of dynamic rubber friction,\
  \bibfield  {journal} {\bibinfo  {journal} {Wear}\ }\textbf {\bibinfo {volume}
  {6}},\ \bibinfo {pages} {375} (\bibinfo {year} {1963})}\BibitemShut {NoStop}%
\bibitem [{\citenamefont {Persson}\ and\ \citenamefont
  {Xu}(2025)}]{persson2025rubber}%
  \BibitemOpen
  \bibfield  {author} {\bibinfo {author} {\bibfnamefont {B.}~\bibnamefont
  {Persson}}\ and\ \bibinfo {author} {\bibfnamefont {R.}~\bibnamefont {Xu}},\
  }\href@noop {} {Rubber friction: Theory, mechanisms, and challenges,\
  \bibfield  {journal} {\bibinfo  {journal} {The Journal of Chemical Physics}\
  }\textbf {\bibinfo {volume} {163}} (\bibinfo {year} {2025})}\BibitemShut
  {NoStop}%
\bibitem [{\citenamefont {Grosch}(1963)}]{grosch1963relation}%
  \BibitemOpen
  \bibfield  {author} {\bibinfo {author} {\bibfnamefont {K.}~\bibnamefont
  {Grosch}},\ }\href@noop {} {The relation between the friction and
  visco-elastic properties of rubber,\ \bibfield  {journal} {\bibinfo
  {journal} {Proceedings of the Royal Society of London. Series A. Mathematical
  and Physical Sciences}\ }\textbf {\bibinfo {volume} {274}},\ \bibinfo {pages}
  {21} (\bibinfo {year} {1963})}\BibitemShut {NoStop}%
\bibitem [{\citenamefont {Delavoipi{\`e}re}\ \emph {et~al.}(2018)\citenamefont
  {Delavoipi{\`e}re}, \citenamefont {Tran}, \citenamefont {Verneuil},
  \citenamefont {Heurtefeu}, \citenamefont {Hui},\ and\ \citenamefont
  {Chateauminois}}]{delavoipiere2018friction}%
  \BibitemOpen
  \bibfield  {author} {\bibinfo {author} {\bibfnamefont {J.}~\bibnamefont
  {Delavoipi{\`e}re}}, \bibinfo {author} {\bibfnamefont {Y.}~\bibnamefont
  {Tran}}, \bibinfo {author} {\bibfnamefont {E.}~\bibnamefont {Verneuil}},
  \bibinfo {author} {\bibfnamefont {B.}~\bibnamefont {Heurtefeu}}, \bibinfo
  {author} {\bibfnamefont {C.~Y.}\ \bibnamefont {Hui}}, \ and\ \bibinfo
  {author} {\bibfnamefont {A.}~\bibnamefont {Chateauminois}},\ }\href@noop {}
  {Friction of poroelastic contacts with thin hydrogel films,\ \bibfield
  {journal} {\bibinfo  {journal} {Langmuir}\ }\textbf {\bibinfo {volume}
  {34}},\ \bibinfo {pages} {9617} (\bibinfo {year} {2018})}\BibitemShut
  {NoStop}%
\bibitem [{\citenamefont {Cuccia}\ \emph {et~al.}(2020)\citenamefont {Cuccia},
  \citenamefont {Pothineni}, \citenamefont {Wu}, \citenamefont
  {M{\'e}ndez~Harper},\ and\ \citenamefont {Burton}}]{cuccia2020pore}%
  \BibitemOpen
  \bibfield  {author} {\bibinfo {author} {\bibfnamefont {N.~L.}\ \bibnamefont
  {Cuccia}}, \bibinfo {author} {\bibfnamefont {S.}~\bibnamefont {Pothineni}},
  \bibinfo {author} {\bibfnamefont {B.}~\bibnamefont {Wu}}, \bibinfo {author}
  {\bibfnamefont {J.}~\bibnamefont {M{\'e}ndez~Harper}}, \ and\ \bibinfo
  {author} {\bibfnamefont {J.~C.}\ \bibnamefont {Burton}},\ }\href@noop {}
  {Pore-size dependence and slow relaxation of hydrogel friction on smooth
  surfaces,\ \bibfield  {journal} {\bibinfo  {journal} {Proceedings of the
  National Academy of Sciences}\ }\textbf {\bibinfo {volume} {117}},\ \bibinfo
  {pages} {11247} (\bibinfo {year} {2020})}\BibitemShut {NoStop}%
\bibitem [{\citenamefont {Venner}\ and\ \citenamefont
  {Lubrecht}(2000)}]{venner2000multi}%
  \BibitemOpen
  \bibfield  {author} {\bibinfo {author} {\bibfnamefont {C.~H.}\ \bibnamefont
  {Venner}}\ and\ \bibinfo {author} {\bibfnamefont {A.~A.}\ \bibnamefont
  {Lubrecht}},\ }\href@noop {} {\emph {\bibinfo {title} {Multi-level methods in
  lubrication}}},\ Vol.~\bibinfo {volume} {37}\ (\bibinfo  {publisher}
  {Elsevier},\ \bibinfo {year} {2000})\BibitemShut {NoStop}%
\bibitem [{\citenamefont {Lugt}\ and\ \citenamefont
  {Morales-Espejel}(2011)}]{lugt2011review}%
  \BibitemOpen
  \bibfield  {author} {\bibinfo {author} {\bibfnamefont {P.~M.}\ \bibnamefont
  {Lugt}}\ and\ \bibinfo {author} {\bibfnamefont {G.~E.}\ \bibnamefont
  {Morales-Espejel}},\ }\href@noop {} {A review of elasto-hydrodynamic
  lubrication theory,\ \bibfield  {journal} {\bibinfo  {journal} {Tribology
  transactions}\ }\textbf {\bibinfo {volume} {54}},\ \bibinfo {pages} {470}
  (\bibinfo {year} {2011})}\BibitemShut {NoStop}%
\bibitem [{\citenamefont {Greenwood}(2020)}]{greenwood2020elastohydrodynamic}%
  \BibitemOpen
  \bibfield  {author} {\bibinfo {author} {\bibfnamefont {J.~A.}\ \bibnamefont
  {Greenwood}},\ }\href@noop {} {Elastohydrodynamic lubrication,\ \bibfield
  {journal} {\bibinfo  {journal} {Lubricants}\ }\textbf {\bibinfo {volume}
  {8}},\ \bibinfo {pages} {51} (\bibinfo {year} {2020})}\BibitemShut {NoStop}%
\bibitem [{\citenamefont {Jacobs}\ \emph {et~al.}(2017)\citenamefont {Jacobs},
  \citenamefont {Junge},\ and\ \citenamefont
  {Pastewka}}]{jacobs2017quantitative}%
  \BibitemOpen
  \bibfield  {author} {\bibinfo {author} {\bibfnamefont {T.~D.}\ \bibnamefont
  {Jacobs}}, \bibinfo {author} {\bibfnamefont {T.}~\bibnamefont {Junge}}, \
  and\ \bibinfo {author} {\bibfnamefont {L.}~\bibnamefont {Pastewka}},\
  }\href@noop {} {Quantitative characterization of surface topography using
  spectral analysis,\ \bibfield  {journal} {\bibinfo  {journal} {Surface
  Topography: Metrology and Properties}\ }\textbf {\bibinfo {volume} {5}},\
  \bibinfo {pages} {013001} (\bibinfo {year} {2017})}\BibitemShut {NoStop}%
\bibitem [{\citenamefont {Ardah}\ \emph {et~al.}(2025)\citenamefont {Ardah},
  \citenamefont {Profito},\ and\ \citenamefont
  {Dini}}]{ardah2025comprehensive}%
  \BibitemOpen
  \bibfield  {author} {\bibinfo {author} {\bibfnamefont {S.}~\bibnamefont
  {Ardah}}, \bibinfo {author} {\bibfnamefont {F.~J.}\ \bibnamefont {Profito}},
  \ and\ \bibinfo {author} {\bibfnamefont {D.}~\bibnamefont {Dini}},\
  }\href@noop {} {A comprehensive review and trends in lubrication modelling,\
  \bibfield  {journal} {\bibinfo  {journal} {Advances in Colloid and Interface
  Science}\ ,\ \bibinfo {pages} {103492}} (\bibinfo {year} {2025})}\BibitemShut
  {NoStop}%
\bibitem [{\citenamefont {Patir}\ and\ \citenamefont
  {Cheng}(1978)}]{patir1978}%
  \BibitemOpen
  \bibfield  {author} {\bibinfo {author} {\bibfnamefont {N.}~\bibnamefont
  {Patir}}\ and\ \bibinfo {author} {\bibfnamefont {H.~S.}\ \bibnamefont
  {Cheng}},\ }\href@noop {} {An Average Flow Model for Determining Effects of
  Three-Dimensional Roughness on Partial Hydrodynamic Lubrication,\ \bibfield
  {journal} {\bibinfo  {journal} {Journal of Lubrication Technology}\ }\textbf
  {\bibinfo {volume} {100}},\ \bibinfo {pages} {12} (\bibinfo {year}
  {1978})}\BibitemShut {NoStop}%
\bibitem [{\citenamefont {Patir}\ and\ \citenamefont
  {Cheng}(1979)}]{patir1979application}%
  \BibitemOpen
  \bibfield  {author} {\bibinfo {author} {\bibfnamefont {N.}~\bibnamefont
  {Patir}}\ and\ \bibinfo {author} {\bibfnamefont {H.}~\bibnamefont {Cheng}},\
  }\href@noop {} {Application of average flow model to lubrication between
  rough sliding surfaces,\ \bibfield  {journal} {\bibinfo  {journal} {Journal
  of Lubrication Technology}\ }\textbf {\bibinfo {volume} {101}},\ \bibinfo
  {pages} {220} (\bibinfo {year} {1979})}\BibitemShut {NoStop}%
\bibitem [{\citenamefont {Tripp}(1983)}]{Tripp1983}%
  \BibitemOpen
  \bibfield  {author} {\bibinfo {author} {\bibfnamefont {J.~H.}\ \bibnamefont
  {Tripp}},\ }\href@noop {} {Surface Roughness Effects in Hydrodynamic
  Lubrication: The Flow Factor Method,\ \bibfield  {journal} {\bibinfo
  {journal} {Journal of Lubrication Technology}\ }\textbf {\bibinfo {volume}
  {105}},\ \bibinfo {pages} {458} (\bibinfo {year} {1983})}\BibitemShut
  {NoStop}%
\bibitem [{\citenamefont {Spikes}(1997)}]{spikes1997mixed}%
  \BibitemOpen
  \bibfield  {author} {\bibinfo {author} {\bibfnamefont {H.}~\bibnamefont
  {Spikes}},\ }\href@noop {} {Mixed lubrication—an overview,\ \bibfield
  {journal} {\bibinfo  {journal} {Lubrication Science}\ }\textbf {\bibinfo
  {volume} {9}},\ \bibinfo {pages} {221} (\bibinfo {year} {1997})}\BibitemShut
  {NoStop}%
\bibitem [{\citenamefont {Bongaerts}\ \emph {et~al.}(2007)\citenamefont
  {Bongaerts}, \citenamefont {Fourtouni},\ and\ \citenamefont
  {Stokes}}]{bongaerts2007soft}%
  \BibitemOpen
  \bibfield  {author} {\bibinfo {author} {\bibfnamefont {J.}~\bibnamefont
  {Bongaerts}}, \bibinfo {author} {\bibfnamefont {K.}~\bibnamefont
  {Fourtouni}}, \ and\ \bibinfo {author} {\bibfnamefont {J.}~\bibnamefont
  {Stokes}},\ }\href@noop {} {Soft-tribology: Lubrication in a compliant
  PDMS--PDMS contact,\ \bibfield  {journal} {\bibinfo  {journal} {Tribology
  International}\ }\textbf {\bibinfo {volume} {40}},\ \bibinfo {pages} {1531}
  (\bibinfo {year} {2007})}\BibitemShut {NoStop}%
\bibitem [{\citenamefont {Petrova}\ \emph {et~al.}(2019)\citenamefont
  {Petrova}, \citenamefont {Weber}, \citenamefont {Allain}, \citenamefont
  {Audebert}, \citenamefont {Venner}, \citenamefont {Brouwer},\ and\
  \citenamefont {Bonn}}]{petrova2019fluorescence}%
  \BibitemOpen
  \bibfield  {author} {\bibinfo {author} {\bibfnamefont {D.}~\bibnamefont
  {Petrova}}, \bibinfo {author} {\bibfnamefont {B.}~\bibnamefont {Weber}},
  \bibinfo {author} {\bibfnamefont {C.}~\bibnamefont {Allain}}, \bibinfo
  {author} {\bibfnamefont {P.}~\bibnamefont {Audebert}}, \bibinfo {author}
  {\bibfnamefont {C.~H.}\ \bibnamefont {Venner}}, \bibinfo {author}
  {\bibfnamefont {A.~M.}\ \bibnamefont {Brouwer}}, \ and\ \bibinfo {author}
  {\bibfnamefont {D.}~\bibnamefont {Bonn}},\ }\href@noop {} {Fluorescence
  microscopy visualization of the roughness-induced transition between
  lubrication regimes,\ \bibfield  {journal} {\bibinfo  {journal} {Science
  Advances}\ }\textbf {\bibinfo {volume} {5}},\ \bibinfo {pages} {eaaw4761}
  (\bibinfo {year} {2019})}\BibitemShut {NoStop}%
\bibitem [{\citenamefont {Sadowski}\ and\ \citenamefont
  {Stupkiewicz}(2019)}]{sadowski2019friction}%
  \BibitemOpen
  \bibfield  {author} {\bibinfo {author} {\bibfnamefont {P.}~\bibnamefont
  {Sadowski}}\ and\ \bibinfo {author} {\bibfnamefont {S.}~\bibnamefont
  {Stupkiewicz}},\ }\href@noop {} {Friction in lubricated soft-on-hard,
  hard-on-soft and soft-on-soft sliding contacts,\ \bibfield  {journal}
  {\bibinfo  {journal} {Tribology International}\ }\textbf {\bibinfo {volume}
  {129}},\ \bibinfo {pages} {246} (\bibinfo {year} {2019})}\BibitemShut
  {NoStop}%
\bibitem [{\citenamefont {Dong}\ \emph {et~al.}(2023)\citenamefont {Dong},
  \citenamefont {Moyle}, \citenamefont {Wu}, \citenamefont {Khripin},
  \citenamefont {Hui},\ and\ \citenamefont {Jagota}}]{dong2023transition}%
  \BibitemOpen
  \bibfield  {author} {\bibinfo {author} {\bibfnamefont {H.}~\bibnamefont
  {Dong}}, \bibinfo {author} {\bibfnamefont {N.}~\bibnamefont {Moyle}},
  \bibinfo {author} {\bibfnamefont {H.}~\bibnamefont {Wu}}, \bibinfo {author}
  {\bibfnamefont {C.~Y.}\ \bibnamefont {Khripin}}, \bibinfo {author}
  {\bibfnamefont {C.-Y.}\ \bibnamefont {Hui}}, \ and\ \bibinfo {author}
  {\bibfnamefont {A.}~\bibnamefont {Jagota}},\ }\href@noop {} {The transition
  from Elasto-Hydrodynamic to Mixed Regimes in Lubricated Friction of Soft
  Solid Surfaces,\ \bibfield  {journal} {\bibinfo  {journal} {Advanced
  Materials}\ ,\ \bibinfo {pages} {2211044}} (\bibinfo {year}
  {2023})}\BibitemShut {NoStop}%
\bibitem [{\citenamefont {Dong}\ \emph {et~al.}(2025)\citenamefont {Dong},
  \citenamefont {Siddiquie}, \citenamefont {Xiao}, \citenamefont {Andrews},
  \citenamefont {Bergman}, \citenamefont {Hui},\ and\ \citenamefont
  {Jagota}}]{dong2025role}%
  \BibitemOpen
  \bibfield  {author} {\bibinfo {author} {\bibfnamefont {H.}~\bibnamefont
  {Dong}}, \bibinfo {author} {\bibfnamefont {R.~Y.}\ \bibnamefont {Siddiquie}},
  \bibinfo {author} {\bibfnamefont {X.}~\bibnamefont {Xiao}}, \bibinfo {author}
  {\bibfnamefont {M.}~\bibnamefont {Andrews}}, \bibinfo {author} {\bibfnamefont
  {B.}~\bibnamefont {Bergman}}, \bibinfo {author} {\bibfnamefont {C.-Y.}\
  \bibnamefont {Hui}}, \ and\ \bibinfo {author} {\bibfnamefont
  {A.}~\bibnamefont {Jagota}},\ }\href@noop {} {Role of wettability, adhesion,
  and instabilities in transitions during lubricated sliding friction,\
  \bibfield  {journal} {\bibinfo  {journal} {Proceedings of the National
  Academy of Sciences}\ }\textbf {\bibinfo {volume} {122}},\ \bibinfo {pages}
  {e2421122122} (\bibinfo {year} {2025})}\BibitemShut {NoStop}%
\bibitem [{\citenamefont {Wu-Bavouzet}\ \emph {et~al.}(2007)\citenamefont
  {Wu-Bavouzet}, \citenamefont {Clain-Burckbuchler}, \citenamefont {Buguin},
  \citenamefont {De~Gennes},\ and\ \citenamefont
  {Brochard-Wyart}}]{wu2007stick}%
  \BibitemOpen
  \bibfield  {author} {\bibinfo {author} {\bibfnamefont {F.}~\bibnamefont
  {Wu-Bavouzet}}, \bibinfo {author} {\bibfnamefont {J.}~\bibnamefont
  {Clain-Burckbuchler}}, \bibinfo {author} {\bibfnamefont {A.}~\bibnamefont
  {Buguin}}, \bibinfo {author} {\bibfnamefont {P.-G.}\ \bibnamefont
  {De~Gennes}}, \ and\ \bibinfo {author} {\bibfnamefont {F.}~\bibnamefont
  {Brochard-Wyart}},\ }\href@noop {} {Stick-slip: wet versus dry,\ \bibfield
  {journal} {\bibinfo  {journal} {The Journal of Adhesion}\ }\textbf {\bibinfo
  {volume} {83}},\ \bibinfo {pages} {761} (\bibinfo {year} {2007})}\BibitemShut
  {NoStop}%
\bibitem [{\citenamefont {Oratis}\ \emph {et~al.}(2025)\citenamefont {Oratis},
  \citenamefont {van~den Berg}, \citenamefont {Bertin},\ and\ \citenamefont
  {Snoeijer}}]{oratis2025viscoelastic}%
  \BibitemOpen
  \bibfield  {author} {\bibinfo {author} {\bibfnamefont {A.~T.}\ \bibnamefont
  {Oratis}}, \bibinfo {author} {\bibfnamefont {K.}~\bibnamefont {van~den
  Berg}}, \bibinfo {author} {\bibfnamefont {V.}~\bibnamefont {Bertin}}, \ and\
  \bibinfo {author} {\bibfnamefont {J.~H.}\ \bibnamefont {Snoeijer}},\
  }\href@noop {} {Viscoelastic lubrication of a submerged cylinder sliding down
  an incline,\ \bibfield  {journal} {\bibinfo  {journal} {Europhysics Letters}\
  }\textbf {\bibinfo {volume} {149}},\ \bibinfo {pages} {63002} (\bibinfo
  {year} {2025})}\BibitemShut {NoStop}%
\bibitem [{\citenamefont {Bonn}\ \emph {et~al.}(2009)\citenamefont {Bonn},
  \citenamefont {Eggers}, \citenamefont {Indekeu}, \citenamefont {Meunier},\
  and\ \citenamefont {Rolley}}]{bonn2009wetting}%
  \BibitemOpen
  \bibfield  {author} {\bibinfo {author} {\bibfnamefont {D.}~\bibnamefont
  {Bonn}}, \bibinfo {author} {\bibfnamefont {J.}~\bibnamefont {Eggers}},
  \bibinfo {author} {\bibfnamefont {J.}~\bibnamefont {Indekeu}}, \bibinfo
  {author} {\bibfnamefont {J.}~\bibnamefont {Meunier}}, \ and\ \bibinfo
  {author} {\bibfnamefont {E.}~\bibnamefont {Rolley}},\ }\href@noop {} {Wetting
  and spreading,\ \bibfield  {journal} {\bibinfo  {journal} {Reviews of modern
  physics}\ }\textbf {\bibinfo {volume} {81}},\ \bibinfo {pages} {739}
  (\bibinfo {year} {2009})}\BibitemShut {NoStop}%
\bibitem [{\citenamefont {Snoeijer}\ and\ \citenamefont
  {Andreotti}(2013)}]{snoeijer2013moving}%
  \BibitemOpen
  \bibfield  {author} {\bibinfo {author} {\bibfnamefont {J.~H.}\ \bibnamefont
  {Snoeijer}}\ and\ \bibinfo {author} {\bibfnamefont {B.}~\bibnamefont
  {Andreotti}},\ }\href@noop {} {Moving contact lines: scales, regimes, and
  dynamical transitions,\ \bibfield  {journal} {\bibinfo  {journal} {Annual
  review of fluid mechanics}\ }\textbf {\bibinfo {volume} {45}},\ \bibinfo
  {pages} {269} (\bibinfo {year} {2013})}\BibitemShut {NoStop}%
\bibitem [{\citenamefont {Kansal}\ \emph {et~al.}(2024)\citenamefont {Kansal},
  \citenamefont {Bertin}, \citenamefont {Datt}, \citenamefont {Eggers},\ and\
  \citenamefont {Snoeijer}}]{kansal2024viscoelastic}%
  \BibitemOpen
  \bibfield  {author} {\bibinfo {author} {\bibfnamefont {M.}~\bibnamefont
  {Kansal}}, \bibinfo {author} {\bibfnamefont {V.}~\bibnamefont {Bertin}},
  \bibinfo {author} {\bibfnamefont {C.}~\bibnamefont {Datt}}, \bibinfo {author}
  {\bibfnamefont {J.}~\bibnamefont {Eggers}}, \ and\ \bibinfo {author}
  {\bibfnamefont {J.~H.}\ \bibnamefont {Snoeijer}},\ }\href@noop {}
  {Viscoelastic wetting: Cox--Voinov theory with normal stress effects,\
  \bibfield  {journal} {\bibinfo  {journal} {Journal of fluid mechanics}\
  }\textbf {\bibinfo {volume} {985}},\ \bibinfo {pages} {A17} (\bibinfo {year}
  {2024})}\BibitemShut {NoStop}%
\bibitem [{\citenamefont {Johnson}\ \emph {et~al.}(1972)\citenamefont
  {Johnson}, \citenamefont {Greenwood},\ and\ \citenamefont
  {Poon}}]{johnson1972simple}%
  \BibitemOpen
  \bibfield  {author} {\bibinfo {author} {\bibfnamefont {K.}~\bibnamefont
  {Johnson}}, \bibinfo {author} {\bibfnamefont {J.}~\bibnamefont {Greenwood}},
  \ and\ \bibinfo {author} {\bibfnamefont {S.}~\bibnamefont {Poon}},\
  }\href@noop {} {A simple theory of asperity contact in elastohydro-dynamic
  lubrication,\ \bibfield  {journal} {\bibinfo  {journal} {Wear}\ }\textbf
  {\bibinfo {volume} {19}},\ \bibinfo {pages} {91} (\bibinfo {year}
  {1972})}\BibitemShut {NoStop}%
\bibitem [{\citenamefont {Persson}\ and\ \citenamefont
  {Scaraggi}(2009)}]{persson2009transition}%
  \BibitemOpen
  \bibfield  {author} {\bibinfo {author} {\bibfnamefont {B.}~\bibnamefont
  {Persson}}\ and\ \bibinfo {author} {\bibfnamefont {M.}~\bibnamefont
  {Scaraggi}},\ }\href@noop {} {On the transition from boundary lubrication to
  hydrodynamic lubrication in soft contacts,\ \bibfield  {journal} {\bibinfo
  {journal} {Journal of Physics: Condensed Matter}\ }\textbf {\bibinfo {volume}
  {21}},\ \bibinfo {pages} {185002} (\bibinfo {year} {2009})}\BibitemShut
  {NoStop}%
\bibitem [{\citenamefont {Persson}\ and\ \citenamefont
  {Scaraggi}(2011)}]{persson2011lubricated}%
  \BibitemOpen
  \bibfield  {author} {\bibinfo {author} {\bibfnamefont {B.}~\bibnamefont
  {Persson}}\ and\ \bibinfo {author} {\bibfnamefont {M.}~\bibnamefont
  {Scaraggi}},\ }\href@noop {} {Lubricated sliding dynamics: flow factors and
  Stribeck curve,\ \bibfield  {journal} {\bibinfo  {journal} {The European
  Physical Journal E}\ }\textbf {\bibinfo {volume} {34}},\ \bibinfo {pages}
  {113} (\bibinfo {year} {2011})}\BibitemShut {NoStop}%
\bibitem [{\citenamefont {Scaraggi}\ \emph {et~al.}(2011)\citenamefont
  {Scaraggi}, \citenamefont {Carbone}, \citenamefont {Persson},\ and\
  \citenamefont {Dini}}]{scaraggi2011lubrication}%
  \BibitemOpen
  \bibfield  {author} {\bibinfo {author} {\bibfnamefont {M.}~\bibnamefont
  {Scaraggi}}, \bibinfo {author} {\bibfnamefont {G.}~\bibnamefont {Carbone}},
  \bibinfo {author} {\bibfnamefont {B.~N.}\ \bibnamefont {Persson}}, \ and\
  \bibinfo {author} {\bibfnamefont {D.}~\bibnamefont {Dini}},\ }\href@noop {}
  {Lubrication in soft rough contacts: A novel homogenized approach. Part
  I-Theory,\ \bibfield  {journal} {\bibinfo  {journal} {Soft Matter}\ }\textbf
  {\bibinfo {volume} {7}},\ \bibinfo {pages} {10395} (\bibinfo {year}
  {2011})}\BibitemShut {NoStop}%
\bibitem [{\citenamefont {Jackson}(1997)}]{jackson1997locally}%
  \BibitemOpen
  \bibfield  {author} {\bibinfo {author} {\bibfnamefont {R.}~\bibnamefont
  {Jackson}},\ }\href@noop {} {Locally averaged equations of motion for a
  mixture of identical spherical particles and a Newtonian fluid,\ \bibfield
  {journal} {\bibinfo  {journal} {Chemical Engineering Science}\ }\textbf
  {\bibinfo {volume} {52}},\ \bibinfo {pages} {2457} (\bibinfo {year}
  {1997})}\BibitemShut {NoStop}%
\bibitem [{\citenamefont {Persson}(2007)}]{persson2007relation}%
  \BibitemOpen
  \bibfield  {author} {\bibinfo {author} {\bibfnamefont {B.}~\bibnamefont
  {Persson}},\ }\href@noop {} {Relation between interfacial separation and
  load: a general theory of contact mechanics,\ \bibfield  {journal} {\bibinfo
  {journal} {Physical Review Letters}\ }\textbf {\bibinfo {volume} {99}},\
  \bibinfo {pages} {125502} (\bibinfo {year} {2007})}\BibitemShut {NoStop}%
\bibitem [{\citenamefont {Snoeijer}\ \emph {et~al.}(2013)\citenamefont
  {Snoeijer}, \citenamefont {Eggers},\ and\ \citenamefont
  {Venner}}]{snoeijer2013similarity}%
  \BibitemOpen
  \bibfield  {author} {\bibinfo {author} {\bibfnamefont {J.~H.}\ \bibnamefont
  {Snoeijer}}, \bibinfo {author} {\bibfnamefont {J.}~\bibnamefont {Eggers}}, \
  and\ \bibinfo {author} {\bibfnamefont {C.~H.}\ \bibnamefont {Venner}},\
  }\href@noop {} {Similarity theory of lubricated Hertzian contacts,\ \bibfield
   {journal} {\bibinfo  {journal} {Physics of Fluids}\ }\textbf {\bibinfo
  {volume} {25}} (\bibinfo {year} {2013})}\BibitemShut {NoStop}%
\bibitem [{\citenamefont {Johnson}(1987)}]{johnson1987contact}%
  \BibitemOpen
  \bibfield  {author} {\bibinfo {author} {\bibfnamefont {K.~L.}\ \bibnamefont
  {Johnson}},\ }\href@noop {} {\emph {\bibinfo {title} {Contact Mechanics}}}\
  (\bibinfo  {publisher} {Cambridge University Press},\ \bibinfo {year}
  {1987})\BibitemShut {NoStop}%
\bibitem [{\citenamefont {Dowson}\ and\ \citenamefont
  {Higginson}(2014)}]{dowson2014elasto}%
  \BibitemOpen
  \bibfield  {author} {\bibinfo {author} {\bibfnamefont {D.}~\bibnamefont
  {Dowson}}\ and\ \bibinfo {author} {\bibfnamefont {G.~R.}\ \bibnamefont
  {Higginson}},\ }\href@noop {} {\emph {\bibinfo {title} {Elasto-hydrodynamic
  lubrication: international series on materials science and technology}}},\
  Vol.~\bibinfo {volume} {23}\ (\bibinfo  {publisher} {Elsevier},\ \bibinfo
  {year} {2014})\BibitemShut {NoStop}%
\bibitem [{\citenamefont {Moes}(1992)}]{moes1992optimum}%
  \BibitemOpen
  \bibfield  {author} {\bibinfo {author} {\bibfnamefont {H.}~\bibnamefont
  {Moes}},\ }\href@noop {} {Optimum similarity analysis with applications to
  elastohydrodynamic lubrication,\ \bibfield  {journal} {\bibinfo  {journal}
  {Wear}\ }\textbf {\bibinfo {volume} {159}},\ \bibinfo {pages} {57} (\bibinfo
  {year} {1992})}\BibitemShut {NoStop}%
\bibitem [{\citenamefont {Essink}\ \emph {et~al.}(2021)\citenamefont {Essink},
  \citenamefont {Pandey}, \citenamefont {Karpitschka}, \citenamefont {Venner},\
  and\ \citenamefont {Snoeijer}}]{essink2021regimes}%
  \BibitemOpen
  \bibfield  {author} {\bibinfo {author} {\bibfnamefont {M.~H.}\ \bibnamefont
  {Essink}}, \bibinfo {author} {\bibfnamefont {A.}~\bibnamefont {Pandey}},
  \bibinfo {author} {\bibfnamefont {S.}~\bibnamefont {Karpitschka}}, \bibinfo
  {author} {\bibfnamefont {C.~H.}\ \bibnamefont {Venner}}, \ and\ \bibinfo
  {author} {\bibfnamefont {J.~H.}\ \bibnamefont {Snoeijer}},\ }\href@noop {}
  {Regimes of soft lubrication,\ \bibfield  {journal} {\bibinfo  {journal}
  {Journal of Fluid Mechanics}\ }\textbf {\bibinfo {volume} {915}},\ \bibinfo
  {pages} {A49} (\bibinfo {year} {2021})}\BibitemShut {NoStop}%
\bibitem [{\citenamefont {Yariv}\ \emph {et~al.}(2024)\citenamefont {Yariv},
  \citenamefont {Brand{\~a}o}, \citenamefont {Wood}, \citenamefont
  {Szafraniec}, \citenamefont {Higgins}, \citenamefont {Bazazi}, \citenamefont
  {Pearce},\ and\ \citenamefont {Stone}}]{yariv2024hydrodynamic}%
  \BibitemOpen
  \bibfield  {author} {\bibinfo {author} {\bibfnamefont {E.}~\bibnamefont
  {Yariv}}, \bibinfo {author} {\bibfnamefont {R.}~\bibnamefont {Brand{\~a}o}},
  \bibinfo {author} {\bibfnamefont {D.~K.}\ \bibnamefont {Wood}}, \bibinfo
  {author} {\bibfnamefont {H.}~\bibnamefont {Szafraniec}}, \bibinfo {author}
  {\bibfnamefont {J.~M.}\ \bibnamefont {Higgins}}, \bibinfo {author}
  {\bibfnamefont {P.}~\bibnamefont {Bazazi}}, \bibinfo {author} {\bibfnamefont
  {P.}~\bibnamefont {Pearce}}, \ and\ \bibinfo {author} {\bibfnamefont {H.~A.}\
  \bibnamefont {Stone}},\ }\href@noop {} {Hydrodynamic interactions between
  rough surfaces,\ \bibfield  {journal} {\bibinfo  {journal} {Physical Review
  Fluids}\ }\textbf {\bibinfo {volume} {9}},\ \bibinfo {pages} {L032301}
  (\bibinfo {year} {2024})}\BibitemShut {NoStop}%
\bibitem [{\citenamefont {Skotheim}\ and\ \citenamefont
  {Mahadevan}(2005)}]{skotheim2005soft}%
  \BibitemOpen
  \bibfield  {author} {\bibinfo {author} {\bibfnamefont {J.}~\bibnamefont
  {Skotheim}}\ and\ \bibinfo {author} {\bibfnamefont {L.}~\bibnamefont
  {Mahadevan}},\ }\href@noop {} {Soft lubrication: The elastohydrodynamics of
  nonconforming and conforming contacts,\ \bibfield  {journal} {\bibinfo
  {journal} {Physics of Fluids}\ }\textbf {\bibinfo {volume} {17}} (\bibinfo
  {year} {2005})}\BibitemShut {NoStop}%
\bibitem [{\citenamefont {Corless}\ \emph {et~al.}(1996)\citenamefont
  {Corless}, \citenamefont {Gonnet}, \citenamefont {Hare}, \citenamefont
  {Jeffrey},\ and\ \citenamefont {Knuth}}]{corless1996lambert}%
  \BibitemOpen
  \bibfield  {author} {\bibinfo {author} {\bibfnamefont {R.~M.}\ \bibnamefont
  {Corless}}, \bibinfo {author} {\bibfnamefont {G.~H.}\ \bibnamefont {Gonnet}},
  \bibinfo {author} {\bibfnamefont {D.~E.}\ \bibnamefont {Hare}}, \bibinfo
  {author} {\bibfnamefont {D.~J.}\ \bibnamefont {Jeffrey}}, \ and\ \bibinfo
  {author} {\bibfnamefont {D.~E.}\ \bibnamefont {Knuth}},\ }\href@noop {} {On
  the Lambert W function,\ \bibfield  {journal} {\bibinfo  {journal} {Advances
  in Computational mathematics}\ }\textbf {\bibinfo {volume} {5}},\ \bibinfo
  {pages} {329} (\bibinfo {year} {1996})}\BibitemShut {NoStop}%
\bibitem [{\citenamefont {Choo}\ \emph {et~al.}(2008)\citenamefont {Choo},
  \citenamefont {Olver}, \citenamefont {Spikes}, \citenamefont {Dumont},\ and\
  \citenamefont {Ioannides}}]{choo2008}%
  \BibitemOpen
  \bibfield  {author} {\bibinfo {author} {\bibfnamefont {J.~W.}\ \bibnamefont
  {Choo}}, \bibinfo {author} {\bibfnamefont {A.~V.}\ \bibnamefont {Olver}},
  \bibinfo {author} {\bibfnamefont {H.~A.}\ \bibnamefont {Spikes}}, \bibinfo
  {author} {\bibfnamefont {M.-L.}\ \bibnamefont {Dumont}}, \ and\ \bibinfo
  {author} {\bibfnamefont {E.}~\bibnamefont {Ioannides}},\ }\href@noop {}
  {Interaction of Asperities on Opposing Surfaces in Thin Film, Mixed
  Elastohydrodynamic Lubrication,\ \bibfield  {journal} {\bibinfo  {journal}
  {Journal of Tribology}\ }\textbf {\bibinfo {volume} {130}},\ \bibinfo {pages}
  {021505} (\bibinfo {year} {2008})}\BibitemShut {NoStop}%
\bibitem [{\citenamefont {Peng}\ \emph
  {et~al.}(2021{\natexlab{a}})\citenamefont {Peng}, \citenamefont {Serfass},
  \citenamefont {Hill},\ and\ \citenamefont {Hsiao}}]{peng2021bending}%
  \BibitemOpen
  \bibfield  {author} {\bibinfo {author} {\bibfnamefont {Y.}~\bibnamefont
  {Peng}}, \bibinfo {author} {\bibfnamefont {C.}~\bibnamefont {Serfass}},
  \bibinfo {author} {\bibfnamefont {C.}~\bibnamefont {Hill}}, \ and\ \bibinfo
  {author} {\bibfnamefont {L.}~\bibnamefont {Hsiao}},\ }\href@noop {} {Bending
  of soft micropatterns in elastohydrodynamic lubrication tribology,\ \bibfield
   {journal} {\bibinfo  {journal} {Experimental Mechanics}\ }\textbf {\bibinfo
  {volume} {61}},\ \bibinfo {pages} {969} (\bibinfo {year}
  {2021}{\natexlab{a}})}\BibitemShut {NoStop}%
\bibitem [{\citenamefont {Peng}\ \emph
  {et~al.}(2021{\natexlab{b}})\citenamefont {Peng}, \citenamefont {Serfass},
  \citenamefont {Kawazoe}, \citenamefont {Shao}, \citenamefont {Gutierrez},
  \citenamefont {Hill}, \citenamefont {Santos}, \citenamefont {Visell},\ and\
  \citenamefont {Hsiao}}]{peng2021elastohydrodynamic}%
  \BibitemOpen
  \bibfield  {author} {\bibinfo {author} {\bibfnamefont {Y.}~\bibnamefont
  {Peng}}, \bibinfo {author} {\bibfnamefont {C.~M.}\ \bibnamefont {Serfass}},
  \bibinfo {author} {\bibfnamefont {A.}~\bibnamefont {Kawazoe}}, \bibinfo
  {author} {\bibfnamefont {Y.}~\bibnamefont {Shao}}, \bibinfo {author}
  {\bibfnamefont {K.}~\bibnamefont {Gutierrez}}, \bibinfo {author}
  {\bibfnamefont {C.~N.}\ \bibnamefont {Hill}}, \bibinfo {author}
  {\bibfnamefont {V.~J.}\ \bibnamefont {Santos}}, \bibinfo {author}
  {\bibfnamefont {Y.}~\bibnamefont {Visell}}, \ and\ \bibinfo {author}
  {\bibfnamefont {L.~C.}\ \bibnamefont {Hsiao}},\ }\href@noop {}
  {Elastohydrodynamic friction of robotic and human fingers on soft
  micropatterned substrates,\ \bibfield  {journal} {\bibinfo  {journal} {Nature
  Materials}\ }\textbf {\bibinfo {volume} {20}},\ \bibinfo {pages} {1707}
  (\bibinfo {year} {2021}{\natexlab{b}})}\BibitemShut {NoStop}%
\bibitem [{\citenamefont {Kargar-Estahbanati}\ and\ \citenamefont
  {Rallabandi}(2025)}]{kargar2025non}%
  \BibitemOpen
  \bibfield  {author} {\bibinfo {author} {\bibfnamefont {A.}~\bibnamefont
  {Kargar-Estahbanati}}\ and\ \bibinfo {author} {\bibfnamefont
  {B.}~\bibnamefont {Rallabandi}},\ }\href@noop {} {Non-monotonic frictional
  behavior in the lubricated sliding of soft patterned surfaces,\ \bibfield
  {journal} {\bibinfo  {journal} {Soft Matter}\ }\textbf {\bibinfo {volume}
  {21}},\ \bibinfo {pages} {448} (\bibinfo {year} {2025})}\BibitemShut
  {NoStop}%
\bibitem [{\citenamefont {Minten}\ and\ \citenamefont
  {Rallabandi}(2026)}]{minten2026hydrodynamic}%
  \BibitemOpen
  \bibfield  {author} {\bibinfo {author} {\bibfnamefont {J.}~\bibnamefont
  {Minten}}\ and\ \bibinfo {author} {\bibfnamefont {B.}~\bibnamefont
  {Rallabandi}},\ }\href@noop {} {Hydrodynamic origin of friction between
  suspended rough particles,\ \bibfield  {journal} {\bibinfo  {journal}
  {Journal of Fluid Mechanics}\ }\textbf {\bibinfo {volume} {1031}},\ \bibinfo
  {pages} {A14} (\bibinfo {year} {2026})}\BibitemShut {NoStop}%
\bibitem [{\citenamefont {Jadhao}\ and\ \citenamefont
  {Robbins}(2019)}]{jadhao2019rheological}%
  \BibitemOpen
  \bibfield  {author} {\bibinfo {author} {\bibfnamefont {V.}~\bibnamefont
  {Jadhao}}\ and\ \bibinfo {author} {\bibfnamefont {M.~O.}\ \bibnamefont
  {Robbins}},\ }\href@noop {} {Rheological properties of liquids under
  conditions of elastohydrodynamic lubrication,\ \bibfield  {journal} {\bibinfo
   {journal} {Tribology Letters}\ }\textbf {\bibinfo {volume} {67}},\ \bibinfo
  {pages} {66} (\bibinfo {year} {2019})}\BibitemShut {NoStop}%
\bibitem [{\citenamefont
  {Israelachvili}(2011)}]{israelachvili2011intermolecular}%
  \BibitemOpen
  \bibfield  {author} {\bibinfo {author} {\bibfnamefont {J.~N.}\ \bibnamefont
  {Israelachvili}},\ }\href@noop {} {\emph {\bibinfo {title} {Intermolecular
  and surface forces}}}\ (\bibinfo  {publisher} {Academic press},\ \bibinfo
  {year} {2011})\BibitemShut {NoStop}%
\bibitem [{\citenamefont {Richards}\ \emph {et~al.}(2023)\citenamefont
  {Richards}, \citenamefont {Warren}, \citenamefont {Hodgson}, \citenamefont
  {Lips},\ and\ \citenamefont {Poon}}]{richards2023anomalous}%
  \BibitemOpen
  \bibfield  {author} {\bibinfo {author} {\bibfnamefont {J.~A.}\ \bibnamefont
  {Richards}}, \bibinfo {author} {\bibfnamefont {P.~B.}\ \bibnamefont
  {Warren}}, \bibinfo {author} {\bibfnamefont {D.~J.}\ \bibnamefont {Hodgson}},
  \bibinfo {author} {\bibfnamefont {A.}~\bibnamefont {Lips}}, \ and\ \bibinfo
  {author} {\bibfnamefont {W.~C.}\ \bibnamefont {Poon}},\ }\href@noop {}
  {Anomalous Scaling for Hydrodynamic Lubrication of Conformal Surfaces,\
  \bibfield  {journal} {\bibinfo  {journal} {arXiv preprint arXiv:2306.17696}\
  } (\bibinfo {year} {2023})}\BibitemShut {NoStop}%
\bibitem [{\citenamefont {Guazzelli}\ and\ \citenamefont
  {Pouliquen}(2018)}]{guazzelli2018rheology}%
  \BibitemOpen
  \bibfield  {author} {\bibinfo {author} {\bibfnamefont {{\'E}.}~\bibnamefont
  {Guazzelli}}\ and\ \bibinfo {author} {\bibfnamefont {O.}~\bibnamefont
  {Pouliquen}},\ }\href@noop {} {Rheology of dense granular suspensions,\
  \bibfield  {journal} {\bibinfo  {journal} {Journal of Fluid Mechanics}\
  }\textbf {\bibinfo {volume} {852}},\ \bibinfo {pages} {P1} (\bibinfo {year}
  {2018})}\BibitemShut {NoStop}%
\bibitem [{\citenamefont {Lobry}\ \emph {et~al.}(2019)\citenamefont {Lobry},
  \citenamefont {Lemaire}, \citenamefont {Blanc}, \citenamefont {Gallier},\
  and\ \citenamefont {Peters}}]{lobry2019shear}%
  \BibitemOpen
  \bibfield  {author} {\bibinfo {author} {\bibfnamefont {L.}~\bibnamefont
  {Lobry}}, \bibinfo {author} {\bibfnamefont {E.}~\bibnamefont {Lemaire}},
  \bibinfo {author} {\bibfnamefont {F.}~\bibnamefont {Blanc}}, \bibinfo
  {author} {\bibfnamefont {S.}~\bibnamefont {Gallier}}, \ and\ \bibinfo
  {author} {\bibfnamefont {F.}~\bibnamefont {Peters}},\ }\href@noop {} {Shear
  thinning in non-Brownian suspensions explained by variable friction between
  particles,\ \bibfield  {journal} {\bibinfo  {journal} {Journal of Fluid
  Mechanics}\ }\textbf {\bibinfo {volume} {860}},\ \bibinfo {pages} {682}
  (\bibinfo {year} {2019})}\BibitemShut {NoStop}%
\bibitem [{\citenamefont {Ness}\ \emph {et~al.}(2022)\citenamefont {Ness},
  \citenamefont {Seto},\ and\ \citenamefont {Mari}}]{ness2022physics}%
  \BibitemOpen
  \bibfield  {author} {\bibinfo {author} {\bibfnamefont {C.}~\bibnamefont
  {Ness}}, \bibinfo {author} {\bibfnamefont {R.}~\bibnamefont {Seto}}, \ and\
  \bibinfo {author} {\bibfnamefont {R.}~\bibnamefont {Mari}},\ }\href@noop {}
  {The physics of dense suspensions,\ \bibfield  {journal} {\bibinfo  {journal}
  {Annual Review of Condensed Matter Physics}\ }\textbf {\bibinfo {volume}
  {13}},\ \bibinfo {pages} {97} (\bibinfo {year} {2022})}\BibitemShut {NoStop}%
\end{thebibliography}
\providecommand{\noopsort}[1]{}\providecommand{\singleletter}[1]{#1}%

\end{document}